\documentclass[aps, prd, groupedaddress, preprint, tightenlines,  eqsecnum, nofootinbib, showpacs]{revtex4}
\usepackage{epsfig}
\usepackage{setspace}
\usepackage{graphicx}
\usepackage{leqno}
\usepackage{cancel}
\usepackage{amsmath}
\usepackage{amsfonts}
\usepackage{amssymb}
\usepackage{enumerate}
\usepackage{wasysym}

\usepackage{listings}

\usepackage{axodraw, float, slashed, graphicx, amssymb, amsmath}
\usepackage[usenames, dvipsnames]{xcolor}
\usepackage{booktabs}

\usepackage[margin=2.2cm, a4paper, includefoot]{geometry}

\def\spa#1.#2{\left\langle#1\,#2\right\rangle}
\def\spb#1.#2{\left[#1\,#2\right]}
\def\eps{\epsilon}
\def\la{\langle}
\def\ra{\rangle}

\def\Aloop{A^{\oneloop}}
\def\BR#1#2{[#1|{K_{abc}}|#2\ra}
\def\BRTTT#1#2{\la#1^+|\slashed{K}_{abc}|#2^+\ra}
\def\NeqEight{\Neq8}
\def\NeqSix{\Neq6}
\def\NeqFour{\Neq4}
\def\NeqOne{\Neq1}

\def\NeqZero{[0]}

\def\loq{=}
\def\pluslo{+{\cal O}(\spa{c}.{d})}

\newcommand{\oneloop}{\text{1-loop}}

\newcommand{\tree}{\text{tree}}
\newcommand{\Neq}[1]{\mathcal{N} = #1}

\newcommand\figref[1]{fig.~\ref{#1}}

\def\be{\begin{equation}}
\def\ee{\end{equation}}

\begin{document}

\hfill\today

\title{Complex Factorisation and Recursion for One-Loop Amplitudes}

\author{Sam D. Alston, David~C.~Dunbar and Warren~B.~Perkins}

\affiliation{
College of Science, \\
Swansea University, \\
Swansea, SA2 8PP, UK\\
\today
}

\begin{abstract}

We consider the factorisation of one-loop amplitudes at complex
kinematic points.  
By  determining the terms that are absent for real
kinematics, 
we can construct a recursive ansatz for
the purely rational pieces of one-loop amplitudes in massless
theories. We illustrate this method by verifying the
Bern {\em et.al.} $n$-point ansatze for the {\em single-minus}
one-loop amplitudes in Yang-Mills theory and by constructing the 
scalar  contribution  to the one-loop five graviton MHV scattering amplitude.

\end{abstract}

\pacs{04.65.+e%
}
\maketitle

\section{Introduction}

In recent years significant progress has been made in the computation
of scattering amplitudes in gauge theories and gravity by utilizing the
analytic properties of these amplitudes~\cite{Eden,Bern:1994zx,Bern:1994cg,Britto:2005fq,Risager:2005vk} .  One ingredient in this
process has been the use of factorisation properties when the momenta
in the amplitude 
have been complexified\footnote{A null momentum can be represented as a
pair of two component spinors $p^\mu =\sigma^\mu_{\alpha\dot\alpha}
\lambda^{\alpha}\bar\lambda^{\dot\alpha}$. For real momenta
$\lambda=\pm\bar\lambda^*$ but for complex momenta $\lambda$ and
$\bar\lambda$ are independent~\cite{Witten:2003nn} .}.    
In particular,
on-shell recursive methods have been very useful in the evaluation of
many massless tree level processes~\cite{Britto:2005fq,Risager:2005vk}.   For
example,  consider shifting
two of the external  momenta
according to:  
\begin{equation}
    \begin{aligned}
        {\overline{\lambda}}_i &\longrightarrow
        \hat{\overline{\lambda}}_i =\overline{\lambda}_i - z
\overline{\lambda}_j
        \\
        { \lambda}_j &\longrightarrow             \hat{           \lambda}_j=  \lambda_j + z
   \lambda_i
    \end{aligned}
        \label{eq:abshift}
\end{equation}
where $z$ is a complex parameter. Providing the shifted amplitude $A(z)$ is
analytic and vanishes at large $\vert z\vert$, then by Cauchy's
theorem,
we may obtain the unshifted function from the residues at the poles in $A(z)$, 
\begin{equation}
A(0)=-\sum_i {\rm Res}\Biggl({A(z)\over z}\Biggr)\Biggr\vert_{z_i}
\label{EQcauchy}
\end{equation}
This is only useful if we can evaluate the residues and hence
an understanding of the  singularity structure of the amplitude is essential. 
At tree level the factorisation is relatively simple:
amplitudes must factorise on multi-particle and collinear poles.
Defining $K^\mu \equiv \sum_{j=i}^{i+r-1} k_j^\mu$,
as $K$ becomes null  the $n$-point tree amplitude
$A_n^\tree$ factorises as
\begin{equation}
A_{n}^{\tree}\ \mathop{\longrightarrow}^{K^2 \rightarrow 0}
\sum_{\lambda} \Biggl[ A_{r+1}^{\tree}  \big(k_i, \ldots,
k_{i+r-1}, K^\lambda\big)  
\, {i \over K^2} \, A_{n-r+1}^{\tree} \big(-K^{-\lambda}, k_{i+r}, \ldots,
    k_{i-1} \big) \Biggr] 
\end{equation}
 where $\lambda$ denotes the helicity of the intermediate state.
Consequently, simple poles in the shifted amplitude $A(z)$ occur at
values of $z$ where $K^2(z)=0$.  Since $k_a+k_b$ is independent of
$z$,  only those $K$'s containing precisely one
of $k_a$ or $k_b$ will be $z$ dependent. When the corresponding
$K^2(z)$ vanishes the residue will be the product of the tree
amplitudes defined at $z=z_i$. Thus the $n$-point tree
amplitude can be expressed in terms of lower point amplitudes:
\begin{equation}
A_n^\tree (0) \; = \; \sum_{i,\lambda} {A^{\tree,\lambda}_{r_i+1}(z_i)
  {i\over K^2}A^{\tree,-\lambda}_{n-r_i+1}(z_i)},
\label{RecursionTree}
\end{equation}
where the summation over $i$ is only over factorisations where the $a$
and $b$ legs are on opposite sides of the pole. This is the on-shell
recursive expression of~\cite{Britto:2005fq}. 

There are several complications in applying these techniques beyond tree level. Firstly,
loop amplitudes can develop higher order singularities for complex momenta .
While these do not block recursion {\it per se}, they do necessitate an understanding of factorisation beyond leading order. 
Suppose a rational function  has 
a double pole so that 
\begin{equation}
        R(z) = \frac{\alpha}{(z - z_i)^2} + \frac{\beta}{(z - z_i)} + \ldots 
\label{eq:polebehaviour}
\end{equation}
then 
\begin{equation}
 {\rm Res}\Biggl({R(z)\over z}\Biggr)\Biggr\vert_{z_i} = - {\alpha \over z_i^2 } +{\beta \over z_i} 
\end{equation}
and
both the leading and sub-leading, or `pole under the
double pole', terms are needed in order to apply recursion as in eq.~\eqref{EQcauchy}.  In general, the structure of the sub-leading
poles is poorly understood. 

Secondly, in general loop amplitudes contain both rational and non-rational pieces. 
One strategy for computing one-loop amplitudes is
to split the amplitude into a {\it cut-constructible} piece and a purely rational piece,
\begin{equation}
A_n= C_n + R_n
\label{splitEQ}
\end{equation}
The $C_n$ may be computed using unitarity techniques~\cite{Bern:1994zx,Bern:1994cg,Bern:1997sc,Britto:2004nc,Forde:2007mi,Dunbar:2009ax}  and the remaining
$R_n$ may then, in principle, be determined recursively via Cauchy's
theorem {\it provided} the singularities of $R_n=A_n-C_n$ are
understood.

A general $n$-point one-loop amplitude in a massless theory such as
gravity or QCD can be expanded in terms of loop momentum integrals, $I_m[P^d(\ell)]$,
where $m$ denotes the number of vertices in the loop and $P^d(\ell)$ is
a polynomial of degree $d$ in the loop momentum $\ell$.
Performing a Passarino-Veltman reduction~\cite{PassVelt}  on the loop momentum
integrals yields an amplitude (to $O(\eps)$ in the dimensional
reduction parameter $\eps$), 
\begin{equation}
 \Aloop_n=\sum_{i}\, c_i\, I_4^{i}
 +\sum_{j}\, d_{j}\, I_3^{j}
 +\sum_{k}\, e_{k} \,   I_2^{k}
+R_n\,,
\label{basisequn}
\end{equation}
where $c_i,d_j,e_k$ and $R_n$ are rational functions and the $I_4$,
$I_3$, and $I_2$ are scalar box, triangle and bubble functions
respectively.  The mathematical form of these integral functions
depends on whether the momenta flowing into a vertex are null
(massless) or not (massive)~\cite{Bern:1993kr}.    
In terms of this basis we can define 
\begin{equation}
C_n =\sum_i c_i I_4^i  +\sum_j d_j I_3^j  +\sum_k e_k I_2^k 
\end{equation}
The coefficients, $c_i,d_j$ and $e_k$, contain a range of singularities that are
not present in the full amplitude. 
Individual coefficients may contain  {\it spurious singularities} of the form $\Delta^{-P}$,  where $\Delta$ is a
Gram-determinant of an integral function, yet the entire amplitude
is finite as $\Delta \to 0$. These singularities in the coefficients can be
of high-order in $\Delta$ : we will encounter a case of $P=5$ in one
of our examples.  These spurious singularities cancel amongst
the terms in $C_n$ and also, crucially, with the rational term $R_n$.
There are also singularities that occur at the same kinematic points as the physical singularities, but are of
higher order. Again cancellations between the terms in $C_n$ and $R_n$ must remove these higher order poles
from the complete amplitude.

Starting from $C_n$, we can view this cancellation constraint as a means of
generating parts of  $R_n$ as higher order poles generate rational {\it descendants} from the terms in $C_n$. To evaluate the residue at a higher order pole
the integral functions must be expanded to a corresponding order and the derivatives in this Taylor expansion eventually yield rational terms. In this way we
obtain rational descendant terms whose origins lie in both the box and
bubble integral contributions to $C_n$.   This does not completely
specify $R_n$ but leaves the unspecified component free of these higher
order singularities.

In the following sections we describe how, by using axial gauge
methods to understand the complex
factorisation, we can apply recursion to the rational parts of one-loop
amplitudes.  Specifically we consider two examples: 
the $n$-point single-minus one-loop Yang-Mills
amplitude $A_n(1^-,2^+,\cdots ,n^+)$ and the five-point
scalar  supergravity amplitude $M_5(1^-,2^-,3^+,4^+,5^+)$. 
The $n$-point amplitude $A_n(1^-,2^+,\cdots ,n^+)$ vanishes at tree level and consequently
is a purely rational one-loop amplitude. As such it has no cut
constructible parts but it does have multiple poles in complex
momentum.  This amplitude was originally computed using off-shell
methods~\cite{Mahlon}. In ref.~\cite{Bern:2005hs}  a form for the sub-leading
singularity was postulated and recursion used to (re)obtain the
$n$-point formulae. Here we will prove, using axial gauge methods, the explicit form of the sub-leading term 
for the shift used in ref.~\cite{Bern:2005hs}.  

The second
example is a case where rational descendants of the integral
functions in $C_n$ contribute to the rational terms.  The example is that of  one-loop five graviton scattering where 
we have a scalar circulating in the loop. This is the last remaining
calculation to complete the five-graviton scattering amplitude.  The
expansion of this amplitude in the form of (\ref{basisequn}) 
is plagued by high order singularities. We obtain $R_5$ recursively and describe 
how these high order singularities generate contributions to $R_5$ in addition to those coming directly from standard and non-standard factorisations.

\section{Complex Factorisation}

For real momenta 
the factorisation of one-loop massless amplitudes is described
in ref.~\cite{BernChalmers},
\begin{equation}\begin{split}
    \label{LoopFact}
    &A_{n}^{\oneloop} \mathop{\longrightarrow}^{K^2 \rightarrow 0}
    \sum_{\lambda=\pm} 
\Biggl[ A_{r+1}^{\oneloop}\big(k_i, \ldots,     k_{i+r-1}, K^\lambda\big) \, {i \over K^2} \,
    A_{n-r+1}^{\tree}\big((-K)^{-\lambda}, k_{i+r}, \ldots,
    k_{i-1}\big) \\ & + A_{r+1}^{\tree}\big(k_i, \ldots, k_{i+r-1},
    K^\lambda\big) {i\over K^2}
    A_{n-r+1}^{\oneloop}\big((-K)^{-\lambda}, k_{i+r}, \ldots,
    k_{i-1}\big) \\
    & + A_{r+1}^{\tree}\big(k_i, \ldots, k_{i+r-1}, K^\lambda\big)
    {i\over K^2} A_{n-r+1}^{\tree}\big((-K)^{-\lambda}, k_{i+r},
    \ldots, k_{i-1}\big) F_n\big(K^2;k_1, \ldots, k_n\big) \Biggr],
  \end{split}\end{equation}
where the one-loop `factorisation function' $F_n$ is
helicity-independent.   This factorisation has single poles in $K^2$. 
We refer to the singularities given in this equation as the standard
factorisations.  Singularities not contained in eq.~\eqref{LoopFact} we refer to as
``non-standard'' factorisations.

For complex momenta we can acquire higher order poles. For a
two-particle pole
$K^2=(k_a+k_b)^2=2k_a\cdot k_b= \spa{a}.{b}\spb{b}.a$.
For real momentum $\spa{a}.{b}=\pm \spb{a}.{b}^*$ and so both vanish
at the pole\footnote{
As usual we are using a spinor helicity formalism with the usual
spinor products $ \spa{j}.{l} \equiv \langle j^- | l^+ \rangle =
\bar{u}_-(k_j) u_+(k_l)$ and $\spb{j}.{l}\equiv \langle j^+ | l^-
\rangle = \bar{u}_+(k_j) u_-(k_l)$. 
In terms of spinors $\spa{a}.{b}=\epsilon_{\alpha\beta}
\lambda_a^\alpha \lambda_b^{\beta}$  and 
 $\spb{a}.{b}=-\epsilon_{\dot\alpha\dot\beta} \bar\lambda_a^{\dot\alpha} \bar\lambda_b^{\dot\beta}$.
We also use  $\BR{i}{j}$ to denote
$\BRTTT{i}{j}$ with $K_{abc}^\mu =k_a^\mu+k_b^\mu+k_c^\mu$ etc. Also
$s_{ab}=(k_a+k_b)^2$, $t_{abc}=(k_a+k_b+k_c)^2$, etc.}.
However for complex momenta we may have $\spa{a}.{b}=0$
but $\spb{a}.{b}\neq 0$.   So terms such as $\spb{a}.b^2/\spa{a}.b^2$
which are finite for real momenta
can have multiple poles for complex momenta.    
These can be interpreted within eq.~(\ref{LoopFact}) as arising from the
three-point one-loop amplitude acquiring a singularity. 
Specifically, the three-point all-plus (or all-minus) one-loop
amplitude has a pole~\cite{Bern:2005hs}
\begin{equation}
A^{\oneloop}_3( K^+, a^+,b^+) = { 1 \over K^2 } V^\oneloop(K^+,a^+,b^+)
\label{ThreePointEq}
\end{equation}
where, for pure Yang--Mills,
\begin{equation}
V^\oneloop(K^+,a^+,b^+) =-{ i \over 48 \pi^2
}\spb{K}.a\spb{a}.b\spb{b}.K   .  
\end{equation}
For real momenta $A^{\oneloop}_3( K^+, a^+,b^+)$ vanishes as $K^2
\longrightarrow 0$ but it can be singular for complex momenta. 
Equation (\ref{ThreePointEq}) specifies the double pole as
$K^2 \longrightarrow 0$ however, as discussed previously, we require
the subleading pole in order to use recursion.

As an example of the structure of the double pole consider the
amplitude with a single minus helicity 
$A_{n}(a^-,b^+,\ldots,n^+)$\footnote{As usual we are considering the
colour-ordered partial amplitudes.}. This amplitude vanishes to all orders in
perturbation theory in a supersymmetric theory and consequently at
tree level in  Yang-Mills.  It is non-vanishing at one-loop level but,
since the tree amplitude vanishes, is entirely rational.  The all-$n$
form was first obtained by Mahlon~\cite{Mahlon} using off-shell recursion~\cite{BerendsGiele}.
In \cite{Bern:2005hs} the complex factorisation of the single-minus one-loop
Yang-Mills amplitude was considered by
applying the  shift of  eq.~(\ref{eq:abshift}) to the
$\bar\lambda$ of the negative helicity leg 
and the $\lambda$ of an adjacent positive helicity leg. 
For this specific case a  
form for the `pole under the double pole'
was proposed.  
Using this and applying complex recursion the following form
for the amplitude was presented:
\begin{equation}
\begin{aligned}
A^{\oneloop}_{n}( & a^-,b^+,\ldots,n^+)  \\
& = A^{\oneloop}_{n-1}(d^+,\ldots,n^+,\hat{a}^-,\hat{K}_{bc}^+)\frac{i}{K^2_{bc}}
A^{\tree}_3(\hat{b}^+,c^+,-\hat{K}_{bc}^-) \\
& + \sum_{i=4}^{n-1} A^{\tree}_{n-i+2} ((i+1)^+,\ldots,n^+,\hat{a}^-,\hat{K}_{b \ldots i}^-)
\frac{i}{K_{b \ldots i}^2} A_i^{\oneloop}(\hat{b}^+,\ldots,i^+,-\hat{K}_{2 \ldots i}^+) \\
& + A^{\tree}_{n-1}(d^+,\ldots,n^+,\hat{a}^-,\hat{K}_{bc}^- ) \frac{i}{(K_{bc}^2)^2}
V^{\oneloop}(\hat{b}^+,c^+,-\hat{K}_{bc}^+) \\
& \quad \times \left( 1 + K_{bc}^2{\cal S}^{(0)}(\hat a, \hat
  K_{bc}^+,d){\cal S}^{(0)}(c, -\hat K_{bc}^-,\hat b) \right)     \; , 
\label{eq:1looponshellrec}
\end{aligned}
\end{equation}
where
\begin{equation}
{\cal S}^{(0)}(a,s^+,b)= {\la ab\ra\over \la as\ra\la sb\ra}
\qquad {\rm and} \qquad
{\cal S}^{(0)}(a,s^-,b)= -{ [ab]\over [as][sb]} \; .
\end{equation}
Expression (\ref{eq:1looponshellrec}) was shown to match that of
Mahlon~\cite{Mahlon}  up to $n=15$. 
This expression has also been justified using gauge Lorentz
invariance~\cite{Vaman:2008rr}.
The form of the subleading pole used to generate
\eqref{eq:1looponshellrec} in terms of soft-factors is only valid for
the particular shift used~\cite{Brandhuber:2007up}.
 In the next section we
provide an explicit constructive derivation of the sub-leading terms based on a diagrammatic analysis using axial gauge rules.

\section{$n$-point Single minus Yang Mills Amplitudes}

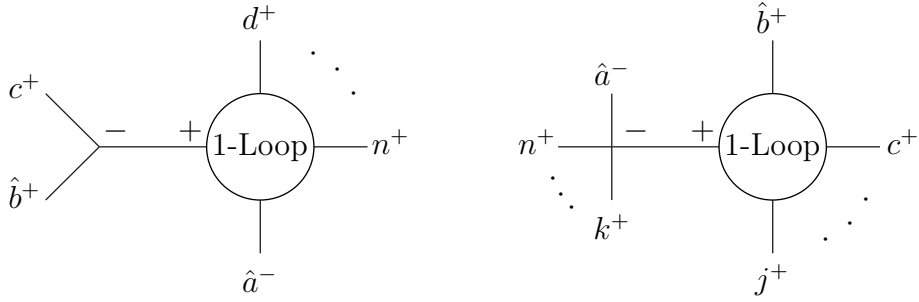
\begin{figure}
  \begin{center}
    {
      \begin{picture}(150,100)(0,-20)
        
        \Line(-20,40)(-40,60)
        \Line(-20,40)(-40,20)
        \Line(-20,40)(80,40)
        \Line(40,0)(40,80)
        \CCirc(40,40){20}{0}{White}
        \Text(40,40)[c]{1-Loop}
        
        \Text( 40,-10)[c]{$\hat a^-$}
        \Text(-48, 23)[c]{$\hat b^+$}
        \Text(-48, 63)[c]{$     c^+$}
        \Text( 40, 89)[c]{$     d^+$}
        \Text( 89, 42)[c]{$     n^+$}
        \Text( 14, 46)[c]{$+$}
        \Text(-14, 46)[c]{$-$}

        \Text(60, 75)[c]{$\cdot$}
        \Text(69, 69)[c]{$\cdot$}
        \Text(75, 60)[c]{$\cdot$}

      \end{picture}
    }
    {
      \begin{picture}(120,100)(-30,-20)
        
        \Line(-20,40)(-20,60)
        \Line(-20,40)(-20,20)
        \Line(-40,40)(80,40)
        \Line(40,0)(40,80)
        \CCirc(40,40){20}{0}{White}
 
        \Text(40,40)[c]{1-Loop}
        \Text( 40,-10)[c]{$     j^+$}
        \Text(-20, 10)[c]{$     k^+$}
        \Text(-48, 42)[c]{$     n^+$}
        \Text(-20, 68)[c]{$\hat a^-$}
        \Text( 40, 89)[c]{$\hat b^+$}
        \Text( 89, 42)[c]{$     c^+$}
        \Text( 14, 46)[c]{$+$}
        \Text(-11, 46)[c]{$-$}
        
        \Text(60,  5)[c]{$\cdot$}
        \Text(69, 11)[c]{$\cdot$}
        \Text(75, 20)[c]{$\cdot$}

        \Text(-35, 17)[c]{$\cdot$}
        \Text(-39, 22)[c]{$\cdot$}
        \Text(-42, 28)[c]{$\cdot$}

      \end{picture}
    }
    \\
    \caption{Standard Factorisations of $A(\hat a^-,\hat b^+,c^+,\ldots, n^+)$.  \label{standfact} }
  \end{center}
\end{figure}

\begin{figure}[htb]\begin{center}
\includegraphics[scale=1.5]{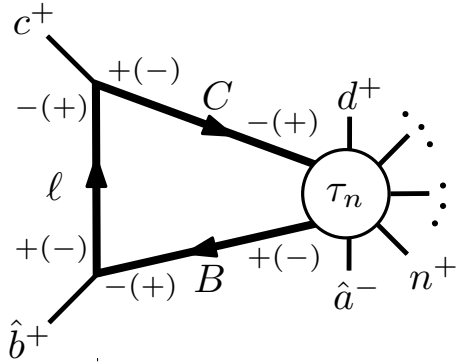}
\caption{The set of diagrams corresponding to the non-standard
  factorisation  as  $\spa{b}.{c}\rightarrow 0$ can be gathered together as
  shown. $\tau_n$ is the set of diagrams with all but two legs
  on-shell.
}
\label{fig:generalintegrand}
\end{center}
\end{figure}

In this section we study the factorisation of the single-minus
Yang-Mills amplitudes $A(a^-,b^+,c^+,\ldots, n^+)$  under a shift of
legs $a$ and $b$ as above.     
The diagrams in \figref{standfact} generate the standard factorisations given in 
eq.(\ref{LoopFact}).
As $\la \hat bc\ra \to
0$ double poles come from diagrams of the form illustrated in
\figref{fig:generalintegrand} where the current $\tau_n$ is the
sum of all possible sub-diagrams. 
To evaluate these diagrams we use axial-gauge
rules~\cite{Schwinn:2005pi}. In this scheme internal off-shell
particles are still labelled by $\pm$ helicity 
and the non-vanishing three-point vertices are the MHV and $\overline{\rm MHV}$ vertices
\begin{equation}
V_3 (1^-,2^-,3^+) = i \frac{
\langle 12 \rangle [ 3 q ]^2 }{ [ 1 q ] [ 2 q ] }
\qquad
V_3 (1^+,2^+,3^-) = i \frac{ [ 21 ] \langle 3 q \rangle^2 }{ \langle
1 q \rangle \langle 2 q \rangle } 
  \label{eq:3ptmhvgoogly}
\end{equation}
where  $q$ is a reference null vector.  For non-null momenta, $P$,  we define
\begin{equation}
|P\ra \equiv P|q] \; ,   \;\   
|P ] \equiv  - {P|q\ra \over  2P\cdot q }
\end{equation}
which corresponds to using 
 {\it q-nullified} momenta $P^\flat$, where 
\begin{equation}
    P^\flat \equiv P - \frac{ P^2 }{ 2 P \cdot q } q .
        \label{eq:offshellK}
\end{equation}

We can simplify this particular computation considerably by setting
$\lambda_q=\lambda_a$ (and leaving $\bar\lambda_q$ arbitrary at this
stage).
With this choice  three-point $\overline{\rm MHV}$ vertices and
four-point MHV vertices involving $a^-$ vanish. Hence the $a^-$ leg must be attached to a three-point MHV
vertex and each diagram contains a single
three-point MHV vertex and $n-1$ three-point $\overline{\rm MHV}$ vertices\footnote{
Consider constructing a one-loop diagram with $n_-$ negative
helicity external legs and $n_+$ positive helicity external legs from $n_3$ three-point
MHV vertices, $\bar n_3$ three point $\overline{\rm MHV}$ vertices and
$n_4$ four-point MHV vertices. Then $n_-=n_3+n_4$ and $n_+=\bar n_3
+n_4$. For our situation with $n_-=1$ there is thus either a single
three or four point MHV vertex.} .

Since the negative helicity leg $a^-$ must be attached to the only  MHV three-point
vertex, the diagrams contributing to the non-standard $\spa{b}.{c}$
poles have the helicity structure shown in \figref{fig:generalintegrand}. 
In this diagram $\tau_n$ is a current, or off-shell tree amplitude.  
We label the off-shell legs (which depend upon loop-momenta) 
\begin{equation}
B = \ell+k_b  \;\;\ 
{\rm and} \;\;\;  C= \ell-k_c \; .
\end{equation}
Note that $B-C=k_b+k_c$. 
As we will see, 
one $\spa{b}.{c}^{-1}$  factor arises from the tree current 
$\tau_n$ and a second from the loop integration, specifically the
region where $\ell$ , $B$ and $C$ are all close to null. 
Throughout, we view the unshifted amplitude as a sum of functions, each of which corresponds to a Feynman 
diagram involving real momenta. In particular the loop momenta are real and where we indicate on a diagram which legs will ultimately be shifted, that shift applies
to the function obtained by evaluating the diagram with real momenta.

The contribution from \figref{fig:generalintegrand}   is then
\def\Icont{{C^{\rm n-s\; f}}}
\begin{equation}
\Icont=\int d^d\ell \frac{ [b|\ell|a\ra [c|\ell|a\ra }{ \la ba \ra \la ca \ra }{\la Ca\ra^2 \over \la Ba\ra^2}
\frac{ \tau_n(a^-,B^-,-C^+,d^+,\ldots,n^+) }{\ell^2 B^2 C^2}
\label{eq:fullloopintegral}
\end{equation}
A factor of $\spa{b}.c^{-1}$ arises from the region of integration where
$\ell^2=0$. Specifically, since $B^2=\ell^2+2\ell\cdot b +b^2 =  \ell^2+2\ell\cdot
b$,  
around $\ell^2=0$,
\begin{equation}
\Icont
\sim \int_0 |\ell|^{d-1} d\ell  { \tau|_{\ell^2=0} \over \ell^2 ( \ell^2 +2\ell\cdot b )
  ( \ell^2 -2\ell\cdot c ) } 
\sim  \int_0 |\ell|^{d-1}  { d \ell \over \ell^2  ( \ell\cdot b ) ( \ell\cdot c )}
\sim { 1\over \spa{b}.c}
\end{equation}

\begin{figure}[htb]\begin{center}
\includegraphics[scale=1]{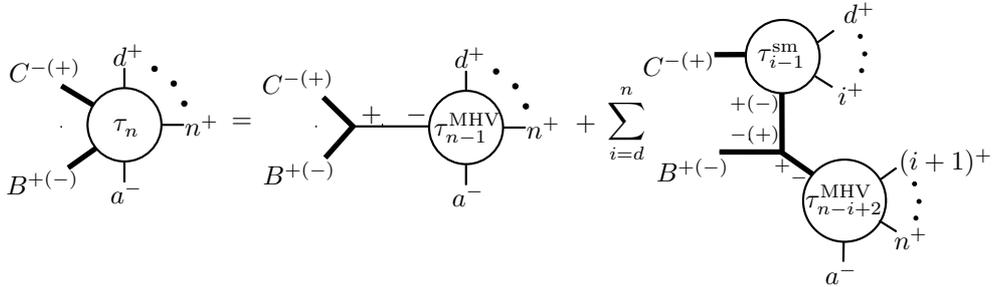}
\caption{An organisation of the diagrams within $\tau_n$.  We have
used the vanishing of the specific single minus current (\ref{EQfirstcurrent}).}
\label{bitsoftau}
\end{center}
\end{figure}

We can expand $\tau_n$ into sub-currents which are either MHV currents
or currents with a single minus as show in \figref{bitsoftau}. 
The first structure in \figref{bitsoftau} we label 
$\tau^{\rm  tri}_n$.   This contains an explicit pole  and generates a further
pole upon integration to give rise to the double
pole contributions. 
The other structures only generate single poles and we label them  $\tau^{\rm  b}_n$.    
The diagrammatic expansion gives both of these contributions in terms
of off-shell MHV tree currents: $\tau^{\rm  MHV}_n$.
We can use the general results, specialised to $\lambda_q=\lambda_a$,
for the currents with one off-shell momentum, $P$, given by~\cite{Kosower:1989xy,Mahlon},
\begin{align}
\tau^{\rm sm}(a^-,P^+,\ldots,n^+) &= 0
\label{EQfirstcurrent} 
 \\
\tau^{\rm sm}(P^-,1^+,\ldots,n^+) &= -
\frac{ P^2 \la P a \ra^2 } { \la a 1 \ra\la 12 \ra \ldots \la n-1,n \ra
  \la n a \ra }
\label{EQsecondcurrent} 
\\
\tau^{\rm MHV}(a^-,P^-,f^+....n^+)&=
{i\la Pa\ra^2P^2\over \la af\ra\la fg\ra\cdots\la (n-1)n\ra\la na\ra}
\Biggl( {[qn] \over [aq] [an]}
- \sum_{l=f}^{n-1} { \la a|P_l l|a\ra \over  P_l^2 P_{l-1}^2}
\Biggr)
\label{offshellmhv}
\end{align}
where $P_l \equiv k_{l+1}+ \cdots +k_n+k_a$. Although the last expression
contains an explicit $P^2$ factor in the numerator, the term in the
sum with $l=f$ contains a $1/P^2$ since $P_{f-1}=-P$ and so survives
in the $P^2\to 0$ limit.  We will need  a simple generalisation of
(\ref{EQsecondcurrent}), 
\begin{equation}
\tau^{\rm sm}(P^{-h},f^{+h}....n^+)=\Biggl({\la fa\ra\over \la
  Pa\ra}\Biggr)^{2-2h}\tau^{\rm sm}(P^-,f^+....n^+) 
\label{SWIlikeB}
\end{equation}
This result follows diagram by diagram in $\tau^{\rm sm}$ as only three-point $\overline{\rm MHV}$  vertices are present and every one that the non-gluonic particle
encounters introduces a factor of
\begin{equation}
\Biggl({[\kappa p_{\rm in} ]\over [\kappa p_{\rm out}]}\Biggr)^{2-2h}
= \Biggl({\la p_{\rm out}a\ra\over \la p_{\rm in}a\ra}\Biggr)^{2-2h}\Biggl({[\kappa|p_{\rm in} |a\ra \over [\kappa|p_{\rm out}|a\ra}\Biggr)^{2-2h}
= \Biggl({\la p_{\rm out}a\ra\over \la p_{\rm in}a\ra}\Biggr)^{2-2h},
\end{equation}
the product of which  gives the factor in (\ref{SWIlikeB}) for each diagram.

For $\tau^{\rm  tri}_n$ we have
\begin{equation}
\begin{aligned}
\tau^{\rm tri}_n &= V_3(P^+,B^+,-C^-) \frac{1}{P^2_{BC}} \tau^{\rm MHV}_{n-1}(a^-, -P^-,d^+,\ldots,n^+) 
\\
&={\la Ca\ra^2\over \la Ba\ra^2}\frac{ \la a | B C | a \ra }{ s_{bc}
}\frac{ \tau^{\rm MHV}_{n-1}(a^-,(b+c)^-,d^+,\ldots,n^+) }{ \la
  P_{b+c} a \ra^2} 
\\ 
\end{aligned}
\label{eq:tautrin}
\end{equation}
This has a very simple dependence on the off-shell momenta $B$ and $C$
and contains an explicit $1/s_{bc}$ factor which, together with the
pole arising from the integration,  produces the double pole. This
term also contributes to the subleading pole.

The remaining contributions to $\tau_n$  arise from the second class
of diagram in \figref{bitsoftau}.  The integrand does not  do not have
an explicit pole as $\spa{b}.c\rightarrow 0$  and only generates a
single pole after integration.  Since this arises at  $C^2=B^2=0$ , we
can  take $C^2=0$ so that the $\tau^{\rm sm}$ structures in
\figref{bitsoftau} have only one massive leg. In this limit 
we can use the formulae of equations \eqref{EQsecondcurrent}  and
\eqref{offshellmhv} for currents with a single massive
leg to obtain (to leading order in $\spa{b}.{c}$)
\begin{align}
  \tau^{b}_n  = \frac{ \la C a \ra^2  }
                      { \la B a \ra^2  } 
\Biggr[ &
      \frac{ \la ca \ra [b|l|a\rangle                                      }
            {   \la c d \ra\la d e\ra \ldots \la n a \ra  [a b]             }
- \sum_{i=d}^{n-1}   
       \frac{ \la ca \ra \la a | B K_i | a \ra                  }
            { \la c d \ra \la d e \ra \ldots \la i a \ra  } 
       \frac{\tau^{\rm MHV}(a^-,-K_i^-,(i+1)^+,\ldots,n^+)}
            { \la K_i  a \ra^2 K_i^2                            }\Biggl]
\label{eq:reducedtau2}
\end{align}
where the $K_i=k_{i+1}+\cdots k_n+k_a$  are fixed by momentum
conservation within the 
$\tau^{\rm MHV}$ structures and we have made use of:
\begin{equation}
{\la Ca \ra \over\la C d\ra} = { \la ca \ra \over \la c d \ra} + {\cal O}(\la bc \ra)
\end{equation}
in the relevant integration region.
In this form we see that
all of the contributions to $\Icont$ involve the same basic integral:
\begin{equation}
\int d^4\ell {[b|\ell|a\ra [c|\ell|a\ra [X|B|a\ra\over \ell^2 B^2 C^2}=
 {i\over 96\pi^2}{\la a|bc|a\ra\over \la bc\ra}[X|2b+c|a\ra 
\label{basicint}
\end{equation}
leading to
\def\Kappa{\tilde\kappa}
\begin{equation}
\begin{aligned}
& \Icont( a^-,b^+,c^+,d^+,e^+,\ldots,n^+)  \\
& ={i\over 96\pi^2} \frac{ [bc] }{ \la bc \ra }\Biggl[ 
{\la a | \beta (b+c) | a\ra\over s_{bc}}
{\tau^{\rm MHV}(a^-,(b+c)^-,d^+\cdots n) \over \la P_{b+c} a\ra^2}
+\frac{ \la ca \ra [b|\beta|a\rangle                                      }
            {   \la c d \ra\la d e\ra \ldots \la n a \ra  [a b]             }
\\
& \hskip40pt
-\sum_{i=d}^{n-1}   
       \frac{ \la ca \ra \la a | \beta K_i | a \ra                  }
            { \la c d \ra \la d e \ra \ldots \la i a \ra  } 
       \frac{\tau^{\rm MHV}(a^-,-K_i^-,(i+1)^+,\ldots,n^+)}
            { \la K_i  a \ra^2 K_i^2                            }\Biggr]
 \\
\end{aligned}
\end{equation}
where $\beta=2b+c$.
Setting $\gamma=-b$ so that $\beta+\gamma=b+c$, we have
\begin{equation}
{\la \gamma a\ra \over \la \gamma d\ra} = {\la b a\ra \over \la b d\ra} {\la cd\ra \over \la cd\ra}={\la c a\ra \over \la c d\ra} +{\cal O}(\la bc \ra) ,
\end{equation}
so to leading order in $\la bc \ra$ we have
\begin{equation}
\begin{aligned}
& \Icont( a^-,b^+,c^+,d^+,e^+,\ldots,n^+)  \\
& ={i\over 96\pi^2} \frac{ [bc] }{ \la bc \ra }\Biggl[ 
{\la a | \beta (b+c) | a\ra\over s_{bc}}
{\tau^{\rm MHV}(a^-,(b+c)^-,d^+\cdots n) \over \la P_{b+c} a\ra^2}
+\frac{ \la \gamma a \ra [b|\beta|a\rangle                                      }
            {   \la \gamma d \ra\la d e\ra \ldots \la n a \ra  [a b]             }
\\
& \hskip40pt
+\sum_{i=d}^{n-1}  {\la a|\beta \kappa_i|a\ra \la  \gamma a\ra^2 \over K_i^2\kappa_i^2} 
               {\tau^{\rm sm}(-\kappa_i^-, \gamma^+,d^+ \cdots i^+)\over \la \kappa_i a\ra^2}
               {\tau^{\rm MHV}(a^-,-K_i^-,(i+1)^+\cdots n) \over \la K_i a\ra^2}\Biggr]
 \\
\label{eq:newrecgeneral}
\end{aligned}
\end{equation}
Again the internal momenta, $\kappa_i$, are specified by momentum conservation within the $\tau^{\rm sm}$  factors.
The quantity in the square brackets is now essentially $\tau_n(a^-,\beta^-,\gamma^+,\cdots,n^+)$. For clarity we can absorb 
the second term into the summation by adopting
an appropriate definition for $\tau^{\rm MHV}(a^-,K_n^-)$.

Using the explicit form for $\tau^{\rm MHV}$ in the $\tau^{\rm tri}$
contribution we have, to leading order,  
\begin{equation}
\begin{aligned}
& \Icont( a^-,b^+,c^+,d^+,e^+,\ldots,n^+)  \\
& ={i\over 96\pi^2} \frac{ [bc] }{ \la bc \ra }\Biggl[ 
{\la a | \beta (b+c) | a\ra\over \la ad\ra\la de\ra\cdots\la (n-1)n\ra\la na\ra}
\Biggl( {[qn] \over [aq] [an]}+{ \la a|(b+c)d|a\ra \over  t_{bcd}s_{bc}}
- \sum_{l=e}^{n-1} { \la a|K_{a,l+1..n}l|a\ra \over  s_{a,l+1..n}s_{a,l..n}}
\Biggr)
\\
& \hskip40pt
+\sum_{i=d}^n  {\la a|\beta \kappa_i|a\ra \la  \gamma a\ra^2 \over K_i^2\kappa_i^2} 
               {\tau^{\rm sm}(-\kappa_i^-, \gamma^+,d^+ \cdots i^+)\over \la \kappa_i a\ra^2}
               {\tau^{\rm MHV}(a^-,-K_i^-,(i+1)^+\cdots n) \over \la K_i a\ra^2}\Biggr]
 \\
\label{eq:newrecgeneralx}
\end{aligned}
\end{equation}

As discussed previously, the contribution to the rational term is 
\begin{equation}
{\rm Res}\biggl( {1\over z}\Icont( \hat a^-,\hat b^+,c^+,d^+,e^+,\ldots,n^+)\biggr)\biggr\vert_{\la \hat bc\ra=0}
\end{equation}
To extract the residue of the double pole term we use:
\begin{equation}
{\rm Res}\biggl(\frac{ [bc] }{ \la \hat bc \ra }{1 \over z t_{\hat bcd}s_{\hat bc}}\bigg)_{\la \hat bc\ra=0}
=
{1 \over  \la bc\ra\la a| (b+c)d|c\ra }\biggl({\la ac\ra\over \la bc\ra} -{[b|c+d|a\ra\la ac\ra\over \la a| (b+c)d|c\ra }\biggr)
\label{dpres} 
\end{equation}
The first term here is precisely the double pole contribution of~\cite{Bern:2005hs}
\begin{align}
 A^{(0)}_{n-1}(d^+,\ldots,n^+,\hat{a}^-,\hat{K}_{bc}^- ) &\frac{i}{(K_{bc}^2)^2}
V^{\oneloop}(\hat{b}^+,c^+,-\hat{K}_{bc}^+)
 = {i\over 96\pi^2}  
{\la ac\ra\over \la bc\ra^2}{\la a | \beta (b+c) | a\ra\over \la cd\ra\la de\ra\cdots\la (n-1)n\ra\la na\ra}
\end{align}
The second term in (\ref{dpres}) contains only a single factor of $\la
bc\ra$ in the denominator and its coefficient is unaffected by the
shift.  We combine this with the single pole pieces of
\eqref{eq:newrecgeneralx} to  write the 
full sub-leading or {\it pole under the pole} (PUP) contribution as:
\def\sop{{\bf\tilde\dagger}}
\begin{align}
{\cal C}^{\rm PUP}={i\over 96\pi^2}&\frac{ [bc] }{ \la bc \ra }
\notag \\&  
\hskip-50pt
\times  \Biggl[
{\la a | \beta (b+c) | a\ra\over \la ad\ra\la de\ra\cdots\la (n-1)n\ra\la na\ra}
\Biggl( {[qn] \over [aq] [an]}
-{ \la a|(b+c)d|a\ra[b|c+d|a\ra\la ac\ra \over  [bc]\la a| (b+c)d|c\ra^2}
- \sum_{l=e}^{n-1} { \la a|K_{a,l+1..n}l|a\ra \over  s_{a,l+1..n}s_{a,l..n}}
\Biggr)
\notag \\ &
+\sum_{i=d}^n  {\la a|\beta \kappa_i|a\ra \la  \gamma a\ra^2 \over K_i^2\kappa_i^2} 
               {\tau^{\rm sm}(-\kappa_i^-, \gamma^+,d^+ \cdots i^+)\over \la \kappa_i a\ra^2}
               {\tau^{\rm MHV}(a^-,-K_i^-,(i+1)^+\cdots n) \over \la K_i a\ra^2}\Biggr]_\sop
\label{pupa}
\end{align}
where $\sop$ denotes that the quantity in square brackets is to be shifted and evaluated at $z=-{\la bc\ra/ \la ac\ra}$. 
The sums in this expression are most of the terms in the expansion of an on-shell MHV amplitude.  We can use the simple interchange properties of the 
Parke-Taylor amplitudes and $\tau^{\rm sm}$ to gather many of these terms
into the finite on-shell MHV amplitude  $\tau^{\rm MHV}(a^-,\beta^-,d^+,\gamma^+,e^+\cdots n^+)$.
From a diagrammatic perspective we have,
\begin{align}&
\tau^{\rm MHV}(a^-,\beta^-,d^+,\gamma^+,e^+\cdots n^+)=
{[d|K_1|a\ra \la \beta a\ra^2\over \la d a\ra} {\tau^{\rm MHV}(a^-,-K_1^-,\gamma^+,e^+\cdots n)\over \la K_1 a\ra^2 K_1^2}
\notag \\ &
+
\la a|\beta\kappa_2|a\ra\la \beta a\ra^2{\tau^{\rm sm}(-\kappa_2^-, d^+, \gamma^+)\over \la \kappa_2 a\ra^2\kappa_2^2}
                                                                                 {\tau^{\rm MHV}(a^-,-K_2^-,e\cdots n) \over \la K_2 a\ra^2 K_2^2}
\notag \\ &
+
\sum_{i=e}^n  \la a|\beta\kappa_i|a\ra\la \beta a\ra^2{\tau^{\rm sm}(-\kappa_i^-, d^+, \gamma^+,e, \cdots i^+)\over \la \kappa_i a\ra^2\kappa_i^2}
                                                       {\tau^{\rm
                                                           MHV}(a^-,-K_i^-,(i+1)^+\cdots
                                                         n)\over  \la       K_i a\ra^2   K_i^2}  \; .
\label{interchaneexpansion}
\end{align}
Interchanging the $\gamma$ and $d$ legs in the $\tau^{\rm sm}$ amplitudes in the final sum introduces a simple factor of the form 
$-({\la ad\ra /\la a\gamma\ra})({\la \gamma e\ra/ \la de\ra})$. We can then replace all but one term of the final sum in (\ref{pupa}) with
$\tau^{\rm MHV}(a^-,\beta^-,d^+,\gamma^+,e^+\cdots n^+)$  and the first two terms in the  expansion (\ref{interchaneexpansion}):  

\begin{align}
{\cal C}^{\rm PUP}=&{i\over 96\pi^2}\frac{ [bc] }{ \la bc \ra }
\notag \\&  
\hskip-50pt
\times  \Biggl[
{\la a | \beta (b+c) | a\ra\over \la ad\ra\la de\ra\cdots\la (n-1)n\ra\la na\ra}
\Biggl( {[qn] \over [aq] [an]}
-{ \la a|(b+c)d|a\ra[b|c+d|a\ra\la ac\ra \over  [bc]\la a| (b+c)d|c\ra^2}
- \sum_{l=e}^{n-1} { \la a|K_{a,l+1..n}l|a\ra \over  s_{a,l+1..n}s_{a,l..n}}
\Biggr)
\notag \\ &
+  {\la a|\beta \kappa_2|a\ra \la  \gamma a\ra^2 \over K_2^2\kappa_2^2} 
               {\tau^{\rm sm}(-\kappa_2^-, \gamma^+,d^+)\over \la \kappa_2 a\ra^2}
               {\tau^{\rm MHV}(a^-,-K_2^-,e^+\cdots n^+) \over \la K_2 a\ra^2}
\notag \\  & 
-{\la ad\ra\la \gamma e\ra \over \la a\gamma\ra\la de\ra}\Biggl(
{[d|K_1|a\ra \la \gamma a\ra^2\over \la d a\ra}  {\tau^{\rm MHV}(a^-,-K_1^-,\gamma^+,e^+\cdots n)\over K_1^2\la K_1 a\ra^2}
\notag \\ &\hskip 100pt
+
{\la a|\beta \kappa_2|a\ra \la  \gamma a\ra^2 \over K_2^2\kappa_2^2} 
               {\tau^{\rm sm}(-\kappa_2^-, d^+,\gamma^+)\over \la \kappa_2 a\ra^2}
               {\tau^{\rm MHV}(a^-,-K_2^-,e^+\cdots n^+) \over \la K_2 a\ra^2}
\notag \\  &\hskip 100pt
-{\la \gamma a\ra^2\over \la \beta a\ra^2}\tau^{\rm MHV}(a^-,\beta^-,d^+,\gamma^+,e^+\cdots n^+)\Biggr)\Biggr]_\sop
\label{pupb}
\end{align}
The shift puts the final $\tau^{\rm MHV}$ on-shell and we can use the Parke-Taylor form for this term. Using (\ref{offshellmhv}) for the off-shell
 $\tau^{\rm MHV}$ factors we see that all of these contain  a common term of
\begin{equation} 
{[qn] \over [aq] [an]} - \sum_{l=e}^{n-1} { \la a|K_{a,l+1..n}l|a\ra \over  s_{a,l+1..n}s_{a,l..n}}
\end{equation}
The overall coefficient of this term vanishes before we apply the shift.
The sum in (\ref{offshellmhv}) for $ \tau^{\rm MHV}(a^-,K_1^-,\gamma^+,e^+\cdots n)$ also contains an $l=\gamma$ term. 
On the pole this cancels with the contribution from the second term in (\ref{pupb}) leaving just the
$\tau^{\rm MHV}(a^-,\beta^-,d^+,\gamma^+,e^+\cdots n^+)$ term,
\begin{align}
{\cal C}^{\rm PUP}=&{i\over 96\pi^2}\frac{ [bc] }{ \la bc \ra }
\times \Biggl[
{\la ad\ra\la \gamma e\ra \over \la a\gamma\ra\la de\ra} {\la \gamma a\ra^2\over \la \beta a\ra^2}\tau^{\rm MHV}(a^-,\beta^-,d^+,\gamma^+,e^+\cdots n^+) \Biggr]_\sop
\label{pupbc}
\end{align}
Performing the shift, under which $\lambda_{\beta},\lambda_{\gamma}\rightarrow
\lambda_c$ at the pole, we obtain
\begin{align}
{\cal C}^{\rm PUP}=&{i\over 96\pi^2}\frac{ [bc] }{ \la bc \ra }
\times
{\la ad\ra\la c e\ra \over \la a c\ra\la de\ra} \tau^{\rm MHV}(a^-,c^-,d^+,c^+,e^+\cdots n^+) 
\label{pupbe}
\end{align}
which exactly reproduces the soft factor of ref.\cite{Bern:2005hs}
given in eq.~\eqref{eq:1looponshellrec}.

\section{Five Graviton Amplitude}

A one-loop graviton scattering amplitude can receive contributions
from a range of particle types circulating in the loop.  We denote the
contribution from a particle of spin-$s$ to the graviton scattering
amplitude by $M^{[s]}_n$ (with $M^{[0]}_n$ representing a real scalar).  
In a supergravity theory there can be contributions from  minimally  coupled matter
multiplets. 
The contributions from the various
supergravity multiplets are~\cite{GravityStringBasedB}\footnote{We
  use the normalisation for the full physical amplitudes
${\cal  M}^\tree=i(\kappa/2)^{n-2} M^\tree, 
{\cal  M}^\oneloop=i(2\pi )^{-2} (\kappa/2)^{n} M^\oneloop$.}
\begin{align}
 M_n^{\Neq8} =&   M_n^{[2]} +8 M_n^{[3/2]}+28 M_n^{[1]} +56 M_n^{[1/2]} +
 70 M_n^{[0]}
\notag\\
M_n^{\Neq6,matter} =&   M_n^{[3/2]}+6 M_n^{[1]} +15 M_n^{[1/2]} +
20 M_n^{[0]}
\notag
\\
  M_n^{\Neq4,matter} =&   M_n^{[1]} +4 M_n^{[1/2]}+ 6M_n^{[0]} 
\notag
\\
 M_n^{\Neq1,matter} =& M_n^{[1/2]} +
 2M_n^{[0]}
\end{align}

These relations can be inverted to obtain a supersymmetric
decomposition of the pure
graviton scattering amplitude, 
\begin{equation}
 M_n^{[2]} =
  M_n^{\Neq8}-8M_n^{\Neq6,matter}+20M_n^{\Neq4,matter}-16M_n^{\Neq1,matter}+2M_n^{[0]} 
\label{EQgravitybits}
\end{equation}

Compared to Yang-Mills theory,  the one-loop amplitudes for graviton
scattering are relatively poorly understood. 
Previously,  only for $n=4$ have all the components of
\eqref{EQgravitybits} been
computed~\cite{GravityStringBasedA,GravityStringBasedB,
  Grisaru:1979re}. 
For $n=5$, the purely rational amplitudes $M_5(+++++)$ and $M_5(-++++)$
only have non-vanishing scalar components which have been computed in
refs.\cite{MaxCalcsB} and \cite{Dunbar:2010xk} respectively.  For the MHV
amplitude $M_5(--+++)$ only the supersymmetric components have been
computed previously: the $\NeqEight$ component in
ref~\cite{MaxCalcsB}, the $\NeqSix$ in \cite{Dunbar:2010fy} and the
$\NeqFour,1$ components in \cite{Dunbar:2011xw,Dunbar:2011dw}. 
In this section we use complex factorisation to obtain the last
remaining component, $M_5^{[0]}$, of the five graviton
scattering amplitude.  

Our starting point is \eqref{basisequn} and its
 the apparently  trivial rewriting:
\begin{equation}
R_n=\sum(Feynman \; Diagrams) - \sum_{i\in \cal C}\, c_i\, I_{4:{\rm trunc}}^{i} -\sum_{k\in \cal E}\, e_{k} \,   I_2^{k},
\label{ratequn}
\end{equation}
where we have absorbed the triangle integral contributions into the
{\it truncated} box contributions~\cite{Bidder:2004vx,Dunbar:2011xw}
and 
identified the amplitude with a sum of Feynman diagrams.
The coefficients of the box and bubble integral functions  in
\eqref{ratequn} are obtained using four dimensional unitarity methods and
we then perform a BCFW~\cite{Britto:2005fq} recursion on \eqref{ratequn} to obtain $R_5$.
Applying a BCFW shift
\begin{equation}
\bar\lambda_1\to \bar\lambda_{\hat 1}=\bar\lambda_1 -z \bar\lambda_3,
\qquad
\lambda_3\to \lambda_{\hat 3}=\lambda_3 +z \lambda_1,
\label{EQspecshift}\end{equation}
we obtain
\begin{equation}
R_5= \sum_{z\neq 0\;  {\rm poles}}{\rm Res}\Biggl({1\over z}\biggl(
{M }(z)   
- \sum_{i\in \cal C}\, \tilde c_i\, I_{4:{\rm trunc}}^{i} -\sum_{k\in \cal E}\, e_{k} \,   I_2^{k} \biggr)\Biggr)
\end{equation} 
subject to the usual caveat about the behaviour of $M(z)$ at large $z$. We denote the contributions arising from the diagrams, boxes and bubbles as $R^{\rm diag}$, $R^{\rm box}$ and $R^{\rm bub}$ respectively, so that,
\begin{equation}
R_5=R_5^{\rm diag} + R_5^{\rm box} + R_5^{\rm bub}.
\label{EQratdeco}
\end{equation}
The  rational {\it descendants} of the box and bubble  contributions are obtained by expanding them around whatever physical or spurious singularities
they contain, while  the poles in the Feynman diagrams correspond to the standard factorisations of the  amplitude and the non-standard factorisations discussed 
previously.

\subsection{Factorisations}

The diagrammatic contribution has two parts: the standard and non-standard factorisations. We therefore set
\begin{equation} 
R_5^{\rm diag}=R_5^{\rm st}+R_5^{\rm ns}
\end{equation}
As there are non-vanishing one-loop scalar single-minus amplitudes, the five-point amplitude has standard factorisations~\eqref{LoopFact} of the form
\begin{align}
M^{\rm tree} (a^-,b^-,-P^+)&{1\over P^2} M^{\rm 1-loop}(P^-,c^+,d^+,e^+),
\notag\\
M^{\rm tree} (a^-,c^+,-P^+)&{1\over P^2} M^{\rm 1-loop}(P^-,b^-,d^+,e^+),
\notag\\
M^{\rm tree} (a^-,c^+,-P^-)&{1\over P^2} M^{\rm 1-loop}(b^-,P^+,d^+,e^+),
\notag\\
M^{\rm tree} (c^+,d^+,-P^-)&{1\over P^2} M^{\rm 1-loop}(a^-,b^-,P^+,e^+).
\end{align}
These contain simple poles so the rational contributions come from the rational parts of the four-point amplitudes. The shift \eqref{EQspecshift} excites six different poles, leading to
\begin{align}
R^{\rm st}_5= & 
             {1\over s_{12}}\Bigl(M^{\rm tree} (\hat1^-,2^-,-P^+)\times R_4(P^-,\hat3^+,4^+,5^+)\Bigr)\Bigr\vert_{s_{\hat 12}=0}
\notag \\   +&
            {1\over s_{14}}\Bigl(M^{\rm tree} (\hat1^-,-P^-,4^+)\times R_4(2^-,P^+,\hat3^+,5^+)\Bigr)\Bigr\vert_{s_{\hat 14}=0}
\notag \\   +&
            {1\over s_{15}}\Bigl(M^{\rm tree} (\hat1^-,-P^-,5^+)\times R_4(2^-,P^+,\hat3^+,4^+)\Bigr)\Bigr\vert_{s_{\hat 15}=0}
 \notag \\   +&
          {1\over s_{23}}\Bigl(M^{\rm tree} (2^-,\hat 3^+,-P^+)\times R_4(\hat 1^-,P^-,4^+,5^+)\Bigr)\Bigr\vert_{s_{2\hat 3}=0}
\notag \\ +&
            {1\over s_{34}}\Bigl(M^{\rm tree} (-P^-,\hat 3^+,4^+)\times R_4(\hat 1^-,2^-,P^+,5^+)\Bigr)\Bigr\vert_{s_{\hat 34}=0}
 \notag \\   +&
           {1\over s_{35}}\Bigl(M^{\rm tree} (-P^-,\hat 3^+,5^+)\times R_4(\hat 1^-,2^-,P^+,4^+)\Bigr)\Bigr\vert_{s_{\hat 35}=0}
\end{align}
where the rational parts of the four-point amplitudes are~\cite{GravityStringBasedB}:
\begin{align}
R_4(a^-,b^+,c^+,d^+)&={1\over 360}\Bigl({s_{ab}s_{ad}\over s_{ac}}\Bigr)^2\Bigl ({\spb{b}.d^2 \over \spb{a}.b\spa{b}.c\spa{c}.d\spb{d}.a}\Bigr)^2\Bigl(s_{ab}^2+s_{ab}s_{ad}+s_{ad}^2\Bigr)
\notag \\ 
R_4(a^-,b^-,c^+,d^+)&={1\over 360 s_{ab}^6}\Bigr({s_{ab}s_{ad}\spa{a}.b^3 \over \spa{b}.c\spa{c}.d\spa{d}.a}\Bigr)^2 
                     \Bigl(2s_{ad}^4+23s_{ac}s_{ad}^3+222s_{ac}^2s_{ad}^2+23s_{ac}^3s_{ad}+2s_{ac}^4\Bigr)
\end{align}

\begin{figure}[H]
 \begin{center}
\begin{picture}(140,100)(0,-20)
\ArrowLine(20,20)(20,60)
\ArrowLine(20,60)(40,40)
\ArrowLine(40,40)(20,20)

\Line(5,10)(20,20)
\Line(5,70)(20,60)
\Line(40,40)(60,50)
\Line(40,40)(60,40)
\Line(40,40)(60,30)

\CCirc(40,40){8}{0}{White}

\Text(40,40)[c]{$\tau$}
\Text(31,24)[l]{${}^{\ell+{c}}$}
\Text(31,54)[l]{${}^{\ell-d}$}
\Text(15,40)[c]{$\ell$}
\Text(-5,70)[c]{$d^+$}
\Text(-5,10)[c]{${c}^+$}
\Text(72,55)[c]{$e^+$}
\Text(72,43)[c]{$a^-$}
\Text(72,25)[c]{${b}^-$}
\end{picture}
\caption{The non-standard factorisation diagram at five-point \label{fiveptnonstandard} }
\end{center}
\end{figure}
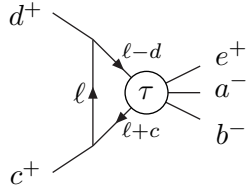

There are also non-standard factorisations.
As discussed in section~3, we expect a complex pole when adjacent
massless legs on a loop become collinear as in
\figref{fiveptnonstandard}.

At 
five-point the tree current, $\tau$, is $\overline{\rm MHV}$ which
does not diverge in the $\la cd\ra\to 0$ limit. We therefore only find $\la cd\ra^{-1}$ poles in this case. 
The region of interest has all three of the illustrated propagators
close to null, so to leading order we can replace $\tau$ by an on-shell tree amplitude
\begin{align}
&\int d^d \ell \biggl({[c|\ell|q\ra\over\la cq\ra} {[d|\ell|q\ra\over\la dq\ra} \biggr)^2 
                       { \tau (e^+,A,B,a^-,b^-)\over \ell^2 A^2
                         B^2}
\rightarrow \int d^d \ell \biggl({[c|\ell|q\ra\over\la cq\ra} {[d|\ell|q\ra\over\la dq\ra} \biggr)^2 
                       {M^{\rm tree}(e^+,A,B,a^-,b^-)\over \ell^2 A^2 B^2},
\end{align}
where $A=\ell+c$, $B=-\ell+d$.  We use the
Kawai-Lewellen-Tye relations~\cite{Kawai:1985xq} to express the
gravity tree amplitude in terms
of Yang-Mills amplitudes,
\begin{align}
M^{\rm tree}(e^+,A,B,a^-,b^-)&=  s_{AB}s_{ab}A^{\rm
  tree}(e^+,A, B,a^-,b^-)A^{\rm tree}(e^+, B, A,b^-,a^-) 
\notag \\ 
&+ 
s_{Aa}s_{Bb}A^{\rm tree}(e^+,A,a^-,B,b^-)A^{\rm tree}(e^+,a^-,A,b^-,B)
\notag \\
= {[eB]^4[eA]^4 \la AB\ra \la ab\ra\over [eA][Ba][be][eB][BA][Ab][ba][ae]}
&+{[eB]^4[eA]^4 \la Aa\ra \la Bb\ra\over [eA][aB][be][ea][aA][Ab][bB][Be]}
\end{align}

In the region of interest $\la AB\ra \sim  0$, so the first term is
negligible, leading to the contribution to $R_5^{\rm ns}$:
\begin{equation}
R_{5:cd}^{\rm ns}\loq\int{ d^d \ell \over \ell^2 A^2 B^2}
{1\over [ae][be]\la cq\ra^2\la dq\ra^2}{[c|\ell|q\ra^2
  [d|\ell|q\ra^2[eB]^3[eA]^3\la Aa\ra \la Bb\ra\over[aB][aA][Ab][bB]} 
\pluslo
\end{equation}

For the five-point amplitude we only need the leading term on the pole which  suggests that the details of the off-shell continuation of $\ell$, $A$ and $B$ 
are not important. 
However the integration region may contain points where one of 
$[aA]$, $[bA]$ , $[aB]$ or $[bB]$ diverge. 
As the only genuine IR divergences in a diagrammatic formulation arise from propagators, we
 rewrite all such factors in the denominator as propagators using:
\begin{equation}
{1\over [xA]}={\la Ax\ra \over [xA]\la Ax\ra}\loq {\la Ax\ra \over
  (A+x)^2}
\pluslo, 
\end{equation}
where the final step involves introducing a sub-leading piece ($A^2$ in this case).  With the labelling
in the figure, the possible propagators involving $A$, $B$, $a$ and $b$ are;
\begin{equation}
{1\over (A+a)^2},\qquad 
{1\over (A+b)^2},\qquad 
{1\over (B-a)^2},\qquad 
{1\over (B-b)^2},
\end{equation}
which fixes the ambiguity with regard to using $(A\pm a)^{-2}$. 
Making use of $\la Ax\ra\la By\ra =\la Ay\ra\la Bx\ra +{\cal O}(\la cd\ra)$ and  setting $\lambda_q=\lambda_a$ we have,
\begin{equation}
R_{5:cd}^{\rm ns} \loq\int{ d^d \ell  \over \ell^2 A^2 B^2}
{1\over [ae][be]\la ca\ra^2\la da\ra^2}{[c|A|b\ra^2 [d|B|b\ra [d|B|a\ra[e|B|a\ra^{3}[e|A|a\ra^{3} 
\over(B-a)^2(B-b)^2(A+a)^2(A+b)^2} 
\pluslo
\end{equation} 
We now apply some leading order reductions:
\begin{equation}
[c|A|b\ra[ba]\la a|A|e]=-2(b\cdot A )[c|Aa|e]+2(a\cdot A)[c|Ab|e] +{\cal O}(\la cd\ra),
\end{equation}
\begin{equation}
[d|B|b\ra[ba]\la a|B|e]=-2(b\cdot B )[d|Ba|e]+2(a\cdot B)[d|Bb|e] +{\cal O}(\la cd\ra),
\end{equation}
leading to,
\begin{equation}
R_{5:cd}^{\rm ns}= I^{[0]}_{a,b}+I^{[0]}_{b,a}-I^{[0]}_{a,a}-I^{[0]}_{b,b}+{\cal O}(\la cd\ra) \; ,
\end{equation}
where
\begin{equation}  
I^{[0]}_{x,y}=
\int{ d^d \ell  \over \ell^2 A^2 B^2}
{1\over [ab]^2[ae][be]\la ca\ra^2\la da\ra^2}{[c|A|b\ra [d|B|a\ra[e|B|a\ra^{2}[e|A|a\ra^{2} [c|Ay|e][d|Bx|e] 
\over(B-x)^2(A+y)^2} \;  . 
\end{equation}
The terms with $x=y$ reduce directly to box integrals using
\begin{equation}
{1\over (B-x)^2(A+x)^2} = {1\over 2P\cdot x}\biggl({1\over (B-x)^2}-{1\over (A+x)^2}\biggr)  +{\cal O}(\la cd\ra).
\end{equation}
Giving $I^{[0]}_{x,x}=I^{[0]}_{x,x:A}+I^{[0]}_{x,x:B}+{\cal O}(\la cd\ra),$ with
\begin{align}
I^{[0]}_{x,x:A}=&
\int{ d^d \ell \over \ell^2 A^2 B^2}
{-1\over [ab]^2[ae][be]\la ca\ra^2\la da\ra^2 2(P\cdot x)}{[c|A|b\ra [d|B|a\ra[e|B|a\ra^{2}[e|A|a\ra^{2} [c|Ax|e][d|Bx|e]
\over (A+x)^2} 
\\
I^{[0]}_{x,x:B}=&
\int{ d^d \ell  \over \ell^2 A^2 B^2}
{1\over [ab]^2[ae][be]\la ca\ra^2\la da\ra^2 2(P\cdot x)}{[c|A|b\ra [d|B|a\ra[e|B|a\ra^{2}[e|A|a\ra^{2} [c|Ax|e][d|Bx|e]
\over(B-x)^2} 
\end{align}
 
For $x\neq y$ successive reductions give
\begin{equation}
I^{[0]}_{x,\bar x}=\sum_{i=1}^5 \bigl(I^{[0]}_{x,\bar
  x:A_i}+I^{[0]}_{x,\bar x:B_i} \bigr) +I^{[0]}_{x,\bar x:P}  +{\cal O}(\la cd\ra),
\end{equation}
where the terms within the sum are the box integrals given explicitly in appendix A and the final term is a cubic pentagon which does not  
contribute to the rational term:
\begin{equation}
I^{[0]}_{x,\bar x:P}=
{[ex]^2[e\bar x]^2 [c\bar x][xd] [\bar x|P|x\ra[x|P|\bar x\ra   \over [ab]^6\la ca\ra^2\la da\ra^2[\bar x x]}
\int{ d^d \ell \over \ell^2 A^2 B^2}
{[c|A|a\ra[d|B|a\ra \la a|P\bar x ABxP|a\ra
\over(B-x)^2(A+\bar x)^2}. 
\end{equation}

\begin{figure}[H]
  \begin{center}
    {
      \begin{picture}(150,144)(0,0)
    \SetOffset(50,72)
        \ArrowLine(-20,-20)(-20, 20)
        \ArrowLine(-20, 20)( 20, 20)
        \ArrowLine( 20, 20)( 20,-20)
        \ArrowLine( 20,-20)(-20,-20)
        
        \Line(-20,-20)(-35,-35)
        \Line(-20, 20)(-35, 35)
        \Line( 20, 20)( 35, 35)
        \Line( 20, -20)( 20, -42)
        \Line( 20, -20)( 42, -20)

        \Text(0,25)[bc]{$B$}
        \Text(0,-25)[tc]{$A$}
        \Text(-25,0)[c]{$\ell$}
        
        \Text(-35, 40)[bc]{$d^+$}
        \Text(-35,-40)[tc]{$c^+$}
        \Text( 43, 40)[bc]{$x^-$}
        \Text( 50, -20)[c]{$\bar x^-$}
        \Text( 20, -48)[c]{$e^+$}

      \end{picture}
    }
    \\
    \caption{The box integral  $I^{[0]}_{x,\bar x:B_5}$  \label{B5Box} }
  \end{center}
\end{figure}
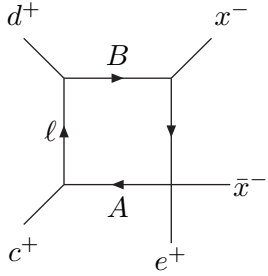

In general, although the non-standard factorisations give both rational and transcendental contributions, we are only interested in the former.
We take $I^{[0]}_{x,\bar x:B_5}$ as an example:
\begin{equation}
I^{[0]}_{x,\bar x:B_5}=
{[ex]^2[e\bar x]^2 [c\bar x][xd] [\bar x|P|x\ra[x|P|\bar x\ra [x|P|a\ra  \over [ab]^6\la ca\ra^2\la da\ra^2[\bar x x]}
\int{ d^d \ell \over \ell^2 A^2 B^2}
{[c|A|a\ra[d|B|a\ra \la x|BP|a\ra
\over(B-x)^2} .
\label{B5eqn}
\end{equation}
We are interested in the rational term generated by the box integral
of \figref{B5Box} with the numerator given in  (\ref{B5eqn}). 

As discussed previously, we expect this box integral to have a $\la cd\ra^{-1}$ singularity.
Dropping the pre-factor, the $\ell$ dependent part of the  integral is
\begin{equation}
I^{\rm bare}_{x,\bar x : B_5}=
\int{ d^d \ell \over \ell^2 A^2 B^2}
{[c|A|a\ra[d|B|a\ra \la x|BP|a\ra
\over(B-x)^2}, 
\label{B5bare}
\end{equation}
Appealing to \eqref{basisequn}, any divergences in this integral will be contained in the cut-constructible pieces so the rational terms of interest here are finite.
Splitting the integration into a four dimensional part and an
$\epsilon$ dimensional part~\cite{Bern:1995db}  we have,  
\begin{equation}
I^{\rm bare}_{x,\bar x :  B_5}= -\epsilon \pi^{-\epsilon}
\Gamma(1-\epsilon)  \times
\int_0^{\infty} { d\mu^2}\int { (\mu^2)^{-1-\epsilon} d^4 \ell \over (\ell^2-\mu^2)( A^2-\mu^2) (B^2-\mu^2)}
{[c|A|a\ra[d|B|a\ra \la x|BP|a\ra
\over(B-x)^2-\mu^2 }, 
\label{B5split}
\end{equation}
where all the momenta are now four dimensional. Parametrising the four dimensional loop momentum by
\begin{equation}
\ell= \alpha k_c +\beta k_d +\gamma \bar\lambda_c\lambda_d +\bar\gamma \bar\lambda_d\lambda_c 
\end{equation}
we see that the first three propagators of eq.~(\ref{B5split}) are of the form 
\begin{equation}
s_{cd}\times\Bigl(f(\alpha,\beta,\vert\gamma\vert) - {\mu^2\over s_{cd}}\Bigr),
\end{equation}
while the final one also contains
\begin{equation}
-2B\cdot x = \alpha s_{cx}+ (\beta-1) s_{dx} +\gamma [xc]\la dx\ra  +\bar\gamma [xd]\la cx\ra=
     \biggl(\alpha+\gamma' \sqrt{s_{dx}\over s_{cx}}\biggr) s_{cx}+ \biggl(\beta-1 +\sqrt{s_{cx}\over s_{dx}} \bar\gamma'\biggr)s_{dx},
\end{equation}
where we have absorbed a pure phase into $\gamma'$. We thus have,
\begin{equation}
(B-x)^2-\mu^2 = s_{cx} \Bigl(F\bigl(\alpha,\beta,\gamma',\bar\gamma',{s_{dx}\over s_{cx}},{s_{cd}\over s_{cx}}\bigl) -{\mu^2\over s_{cx} }\Bigr)
\end{equation}
The first three propagators  of eq.~(\ref{B5split}) combine with the Jacobian to give the required $\la cd\ra^{-1}$ pole while the fourth must introduce a spurious pole: the rational term does 
not involve $\mu$,
so we have an undetermined function depending on ${s_{dx}/ s_{cx}}$ and ${s_{cd}/ s_{cx}}$. As written we have introduced
 a spurious $[cx]$ factor in the denominator, while this could be cancelled by either a ${s_{cx}/ s_{dx}}$ or ${s_{cx}/ s_{cd}}$ factor, either of these would 
introduce a different
spurious factor ( $[dx]$ or $\la cd\ra$ respectively). The rational term can therefore be no more than what is required to cancel the spurious poles generated by the 
cut-constructible pieces
of the integral.

As we are interested in the $\la cd\ra\to 0$ limit, there are two cuts we might consider: $\{d,x\}\{\bar x,e,c\}$ and $\{\bar x,e\}\{c,d,x\}$. 
On spurious poles the coefficients of
pairs of bubbles coincide (up to a sign) so that the logarithms in the
integral functions cancel leaving a  rational descendant. Therefore we only need to calculate one bubble coefficient,
 with the $\{d,x\}\{\bar x,e,c\}$  cut being
the most natural.  A direct parametrization  yields a purely rational
cut integral, so there are no box or triangle contributions to this cut and to 
leading order in $\la cd\ra$ the bubble
coefficient generated by the bare integral is 
\def\xx{x,\bar x :}\begin{equation}
\hskip-70pt
 c^{\rm bub,bare}_{\xx B_5}= 
{1\over \la cd\ra}
\Biggl({[cd]\over [xc]}\Biggr)^{2} {1\over 2}  \la ad\ra^2
\la a|P|x]   
\end{equation}
Here the $[xc]^{-2}$ factor is a remnant of the spurious singularity $[c|d+x|c\ra$ in the $\la cd\ra\to 0$ limit.
Thanks to the $[C|A|x\ra$ factor in the numerator the  order of
this spurious pole is one less than the loop momentum power count.

The rational descendants of the bubbles are obtained by multiplying
the bubble coefficient by the expansion of the difference of the two
integral functions:
\begin{equation}
I_2[s_{dx}]-I_2[s_{e\bar x} ]=
\log\bigl({s_{e\bar x}\over s_{dx}}\bigr) \loq\log\bigl({s_{cx}+s_{dx}\over s_{dx}}\bigr) 
\pluslo 
= {s_{cx}\over s_{dx}} -{1\over 2}{s_{cx}^2\over s_{dx}^2} +{1\over
  3}{s_{cx}^3\over s_{dx}^3}+\ldots 
\end{equation} 
A term involving $[xc]^{-1}$ thus produces a finite rational descendant which need not be cancelled by the rational term. However terms with higher order poles 
would contribute to a spurious
singularity and so must be cancelled. We can read off the appropriate rational term by taking all the terms from the expansion that 
leave at least one factor of $1/[xc]$ 
uncancelled.  
So that this procedure does not introduce other spurious poles we must utilize any factors of $[cx]$ or $[dx]$ present in the numerator of the bare bubble coefficient 
(in particular for
those bare bubble coefficients containing $[ex]$ factors we use
$[cd][ex]= [ed][cx]+[ce][dx]$).   To leading order we can also exploit the $\la ad\ra^n$ factor using:
\begin{equation}
{s_{cx}\over s_{dx}}= { [cx]\la xc\ra \la da\ra \over [dx]\la xd\ra \la da\ra} 
\loq { [cx]\la xd\ra \la ca\ra \over [dx]\la xd\ra \la da\ra} \pluslo ={ [cx] \la ca\ra \over [dx] \la da\ra}  \pluslo
\end{equation}
For the $B_5$ contribution we  note that
\begin{equation}
\hskip-70pt
 c^{\rm bub,bare}_{\xx B_5}= 
{1\over \la cd\ra}
\Biggl({[cd]\over [xc]}\Biggr)^{2} {1\over 2}  \la ad\ra^2
\la a|d|x] +{\cal O}([xc]^{-1})   
\end{equation}
and obtain the rational term,
\begin{equation}
{\cal R}^{\rm bare}_{\xx B_5}(a,b,c,d,e)=\Biggl({1\over \la cd\ra}
\Biggl({[cd]\over [xc]}\Biggr)^{2} {1\over 2}  \la ad\ra^2
\la a|d|x]\Biggr) { [cx] \la ca\ra \over [dx] \la da\ra} = {1\over 2}{[cd]^2\la da\ra^2  \la ca\ra \over \la cd\ra[xc]}
\end{equation}
The rational pieces of the other box integrals can be obtained using the same procedure, giving the rational term
\begin{align}
{\cal R}(a,b,c,d,e)=  & {\cal R}_{a,a:A}(a,b,c,d,e)+{\cal R}_{b,b:A}(a,b,c,d,e)+{\cal R}_{a,a:B}(a,b,c,d,e)+{\cal R}_{b,b:B}(a,b,c,d,e)
                    \notag\\ &+\sum_{i=1}^5 \Bigl({\cal R}_{a,b:A_i}(a,b,c,d,e)+{\cal R}_{b,a:A_i}(a,b,c,d,e)\Bigr)
                    \notag\\ &+\sum_{i=1}^5 \Bigl({\cal R}_{a,b:B_i}(a,b,c,d,e)+{\cal R}_{b,a:B_i}(a,b,c,d,e)\Bigr)
\end{align}
where ${\cal R}_{\xx A_i}$ denotes the full rational term generated by
the box integral $I_{\xx A_i}$ (i.e. the bare integral multiplied by its pre-factor).
There are only simple poles in these contributions so we find
\begin{equation}
R^{ns}_5= {1\over \spa{3}.4}\Bigl({\cal R}(\hat 1,2,\hat 3,4,5)\spa{\hat 3}.4\Bigr)\Bigl\vert_{\spa{\hat 3}.4\to 0}
         +{1\over \spa{3}.5}\Bigl({\cal R}(\hat 1,2,\hat 3,5,4)\spa{\hat 3}.5\Bigr)\Bigl\vert_{\spa{\hat 3}.5\to 0}
\end{equation}

\subsection{Box Contribution}

There is only one type of box at five-point and their coefficients are readily evaluated using quadruple cuts \cite{Britto:2004nc}:
\begin{equation}
c^{\rm scalar}_{\rm box}(c^+,a^-,d^+,\{b^-,e^+\})=
 {(-1)^5\over 2}    {  \la da\ra^4 \la ca\ra^4 [ac]^2[ad]^2\over   \la cd\ra^8}
            {\spb{b}.{e}  \spa{b}.c^3 \spa{b}.d^3\over  \spa{b}.e \spa{c}.{e}\spa{d}.e 
    } \; .
\end{equation}
Around the $\la cd\ra = 0$ pole the truncated one-mass box integral function has the expansion,
\begin{equation} 
I_{4:{\rm trunc}}^{1m}(s,t,u,m)=u\bigl( f_s\log(s)+f_t\log(t) +f_m\log(m)\bigr) +u^2 f_r,
\end{equation}
where 
\begin{equation}
f_x=\sum_{j=0}^{\infty} u^j f_{xj}(s,t,m),
\end{equation}
the $f_{xj}$ are rational functions and for this box $u=s_{cd}$, $s=s_{ac}$, $t=s_{ad}$ and $m=s_{be}$.
Given the $\la cd\ra^{-8}$ singularity in the box coefficient, on the $\la cd\ra =0$ pole the box contributions produce logarithmic and rational descendants 
with leading singularities $\la cd\ra^{-7}$ and $\la cd\ra^{-6}$ respectively.
The logarithmic  descendants combine with the bubble contributions to leave $\la cd\ra^{-1}$ singularities in the effective coefficients of the logarithms. 
The multiple poles in the rational descendants cancel against terms in $R_5$.  

The full rational descendant of this box contribution is
\begin{equation}
{\cal D}_{\rm box}(a,b,c,d,e)=c^{\rm scalar}_{\rm box}(c^+,a^-,d^+,\{b^-,e^+\}) \times {\cal R}^7(s_{ac},s_{ad},s_{cd},s_{be}),
\end{equation}
where, since $c^{\rm scalar}_{\rm box}$ has poles of order eight, ${\cal R}^7$ is
the  expansion of $u^2 f_r$ truncated to seventh order in $u$,
\begin{align}
{\cal R}^7&(s,t,u,m)              =  {u^2\over s^2 t^2} 
                                   -{1\over 3}{u^3 m\over s^3 t^3} 
                                   +{1\over 12}{u^4(2s^2+5st+2t^2) \over s^4 t^4} 
                                   -{1\over 30}{u^5 m(3s^2-2st+3t^2) \over s^5 t^5}
\\ \notag &
                                   +{1\over 720}{u^6 (48s^4+84s^3t-40s^2t^2+84st^3+48t^4) \over s^6 t^6} 
                                   -{1\over 210}{u^7 m(10s^4-8s^3t+9s^2t^2-8st^3+10t^4) \over s^7 t^7}
\end{align}

The shift {\it excites} the $\la cd\ra =0$ type pole in four of the six box contributions, leading to
\begin{align}
R_5^{\rm box}=\;\;\; & {\rm Res}_{\la \hat 34\ra=0}\Biggl({1\over z} \Bigl({\cal D}_{\rm box}(\hat1,2,\hat3,4,5)+{\cal D}_{\rm box}(2,\hat1,\hat3,4,5)\Bigr)\Biggr)  \\ \notag &
                       +{\rm Res}_{\la \hat 35\ra=0}\Biggl({1\over z} \Bigl({\cal D}_{\rm box}(\hat1,2,\hat3,5,4)+{\cal D}_{\rm box}(2,\hat1,\hat3,5,4)\Bigr)\Biggr)
\end{align}

\subsection{Bubble Contribution}

The bubble coefficients can be obtained from two particle cuts using canonical forms~\cite{Dunbar:2009ax}.
There is a single type of bubble with coefficient  $c^{\rm
  scalar}_{\rm bub}(\{a^-,c^+\},\{b^-,d^+,e^+\})$. The general form of
the bubble coefficients is given in appendix~\ref{bubcoeapp}.

The bubble coefficients contain poles of the form $\la cd\ra^{-7}$. On the $\la cd\ra \to 0$ poles these terms are precisely those required to cancel the 
multiple poles in the coefficients of the logarithmic descendants of the boxes. 
The bubble coefficients also contain
 spurious poles of the form $[d|a+c|d\ra^{-5}$ (two powers worse than the $\NeqOne$ case). On these poles pairs of bubble contributions
combine to produce rational descendants with $[d|a+c|d\ra^{-4}$ spurious poles. The singular pieces of these rational descendants  are
\begin{equation}
{\cal D}_{\rm bub}(1,2,3,4,5)= \Bigl({\cal S}(1,2,3,4,5)+{\cal P}_{\{3,4,5\}}\Bigr)+\Bigl({\cal S}(2,1,3,4,5)+{\cal P}_{\{3,4,5\}}\Bigr)
\end{equation}
where ${\cal P}_{\{3,4,5\}}$ represents permutations of the positive helicity legs and
\begin{align}
\hskip-30pt
{\cal S}&(a,b,c,d,e)= 
{ \spa{a}.{c}^4 \spa{b}.d^6 \spb{b}.e s_{cd}^3\over 60\spa{b}.e\spa{c}.d^8\spa{d}.e^2 [d|a+c|d\ra} 
 \notag \\ 
\times\Biggl( &
       {s_{cd}^3\over [d|a+c|d\ra^3}
       -{1\over 2}{s_{cd}^2\over [d|a+c|d\ra^2}
            \biggl[  {s_{cd}\over s_{ac}} +3 \biggl(1+3{\spa{c}.b\spa{a}.d\over \spa{a}.c\spa{b}.d}-{\spa{c}.e\spa{a}.d\over \spa{c}.a\spa{d}.e} \biggr) 
            \biggr]
 \notag \\
       &+ {1\over 12}{s_{cd}\over  [d|a+c|d\ra}
            \biggl[  4{s_{cd}^2\over s_{ac}^2} +9{s_{cd}\over s_{ac}}\biggl(1+3{\spa{c}.b\spa{a}.d\over\spa{a}.c\spa{b}.d}-{\spa{c}.e\spa{a}.d\over\spa{c}.a\spa{d}.e}\biggr)
 \notag \\
              &\hskip 90pt +30\biggl( 3{\spa{c}.b^2\spa{a}.d^2\over\spa{a}.c^2\spa{b}.d^2}
                     -3{\spa{c}.b\spa{a}.d\over\spa{a}.c\spa{b}.d}{\spa{c}.e\spa{a}.d\over\spa{c}.a\spa{d}.e}
                       +{\spa{c}.e^2\spa{a}.d^2\over\spa{c}.a^2\spa{d}.e^2}    
                   \biggr)
            \biggr]   
 \notag \\                        
         +&
               {1\over 2}{s_{cd}^2\over s_{ac}^2}\biggl(-1-3{\spa{c}.b\spa{a}.d\over\spa{a}.c\spa{b}.d}\biggr)  
                 -{15\over 4}{s_{cd}\over s_{ac}}{\spa{a}.d^2\spa{c}.b\spa{c}.d\spa{b}.e\over\spa{a}.c^2\spa{b}.d^2\spa{d}.e} 
                 +10{\spa{c}.e\spa{a}.d\over\spa{a}.c\spa{d}.e        }
 \notag \\
          &  +5
           \biggl[                                                               
             {\spa{c}.d^2\over\spa{a}.c^2}\biggl(  3{\spa{a}.b^2\over \spa{b}.d^2} + 4{\spa{a}.e^2\over\spa{d}.e^2} +9{\spa{a}.b\over\spa{b}.d}{\spa{a}.e\over\spa{d}.e} \biggr )    
             -{\spa{c}.d^3\over\spa{a}.c^3}{\spa{a}.b^2\over\spa{b}.d^2}\biggl( {\spa{a}.b\over\spa{b}.d}+3{\spa{a}.e\over\spa{d}.e} \biggr)         
            \biggr]             
       \Biggr)                       
\end{align} 
 
The shift excites four distinct spurious poles: $[\hat 3|\hat 1+4|\hat 3\ra=0$, $[\hat 3|\hat 1+5|\hat 3\ra=0$, $[4|\hat 1+5|4\ra=0$ and $[5|\hat 1+4|5\ra=0$ 
and we have
\begin{align}
R_5^{\rm bub}(1,2,3,4,5)=\;\;\; & {\rm Res}_{[\hat 3|\hat 1+4|\hat 3\ra=0}\Biggl({1\over z} {\cal D}_{\rm bub}(\hat1,2,\hat3,4,5)\Biggr)
                               +{\rm Res}_{[\hat 3|\hat 1+5|\hat 3\ra=0}\Biggl({1\over z} {\cal D}_{\rm bub}(\hat1,2,\hat3,4,5)\Biggr)   \notag\\ &
                               +{\rm Res}_{[4|\hat 1+5|4\ra=0}\Biggl({1\over z} {\cal D}_{\rm bub}(\hat1,2,\hat3,4,5)\Biggr)
                               +{\rm Res}_{[5|\hat 1+4|5\ra=0}\Biggl({1\over z} {\cal D}_{\rm bub}(\hat1,2,\hat3,4,5)\Biggr)
\end{align}
The residues of all terms are readily extracted using 
{\tt  Mathematica}. This completes the computation of $R_5$. 
Technically, the validity of the above result relies upon, {\it a  priori}, the vanishing of $R(z)$ for large $z$ for which no general
theorems are available. Our expression satisfies several
non-trivial consistency conditions which, experience suggests, make it
almost certainly correct.  
In particular, the
resulting expression for the amplitude has the correct symmetries, is free from spurious
poles, has the 
correct soft and collinear limits and has the correct complex
factorisation. 
The {\tt Mathematica} expression
for $R_5$ is available at 
\url{http://pyweb.swan.ac.uk/~dunbar/graviton.html}.

Our results are consistent with the
suggestion that gravity perturbation theory has a significantly softer
ultra-violet behaviour
at one-loop
than traditionally expected~\cite{BjerrumBohr:2006yw,Bern:2007xj,Dunbar:2010fy}. 
 If a Feynman diagram has $n$-points
in the loop and has a loop momentum polynomial of degree $m$ then
bubble coefficients have $(a\cdot P)$ spurious singularities of the form
\begin{equation}
\sim 
{ 1\over ( a \cdot P)^{m-n+3}   }
\end{equation}
and the rational terms have one power less. 
For gravity the traditional expectation of  $m=2n$ leads to 
$(a\cdot P)^{-n-2}$.  However, our explicit calculations indicate a softer
behaviour.  The bubble coefficients~(\ref{allAform}) have explicit
$(a\cdot P)^{-6}$ singularities arising from the $H^4_{2x}$ terms.
For the five-point case the leading singularity vanishes leaving
$(a\cdot P)^{-5}$ singularities - and consequently only $(a\cdot
P)^{-4}$ singularities in the rational term.  These singularities
are thus consistent with an effective power count of $m=n+4$.
 This is in agreement with the expectations of~\cite{Bern:2007xj,Dunbar:2010fy}. 

\section{Conclusions}

Recursive techniques based on the factorisation of rational tree
amplitudes have proved of great utility. 
In extending these methods to one-loop amplitudes one faces obstacles of
various types: in general one-loop amplitudes contain both  rational and
transcendental functions along with  a plethora of spurious poles and higher
order physical poles. 
The transcendental pieces of one-loop amplitudes are readily obtained using
four dimensional unitarity techniques. 
When the coefficients of these transcendental functions contain high order
poles these contributions give rational descendants which must be accounted for when
computing the remaining rational terms recursively.  
Further, using recursive techniques for the computation of one-loop
amplitudes requires an understanding of the singularities of
these amplitudes when extended to complex momenta. 
For complex momenta there are ``non-standard'' factorisations
which must be accounted for.
 
In this article we have demonstrated how  axial gauge techniques may be
used to determine the non-standard factorisations. 
 Axial gauge techniques  provide a
natural method for examining complex structures since they preserve
the language and structure of on-shell amplitudes, however, like usual
Feynman diagram techniques, they can prove rather cumbersome. 
Nonetheless,  we have used these techniques to A) prove the conjectured sub-leading
pole in the single-minus Yang-Mills amplitude, B) compute the rational
terms of the five graviton
MHV amplitude analytically. This completes the calculation of the
five graviton scattering amplitude.  The expression we obtain is not
particularly simple but will provide a significant target for
alternative techniques. 
Our results are consistent with the
suggestion that gravity perturbation theory has a significantly softer UV behaviour
than traditionally expected.

The work of S. Alston was supported by a STFC studentship.

\appendix

\section{Box Integrals Contributing To The Non-standard Factorisations} 

The non-standard factorisation term $I^{[0]}_{x,\bar x}$ can be expressed as a sum of 
box integrals:
\begin{equation}
I^{[0]}_{x,\bar x}=\sum_{i=1}^5 \bigl(I^{[0]}_{x,\bar
  x:A_i}+I^{[0]}_{x,\bar x:B_i} \bigr) +I^{[0]}_{x,\bar x:P} +{\cal O}
(\spa{c}.{d} ) \; .
\end{equation}
Using the $\ell$-dependent variables $A=\ell+c$, $B=-\ell+d$,  the
integrals required are 
\begin{equation}\hskip-30pt
I^{[0]}_{x,\bar x:A_1}=
{[e\bar x] \over [ab]^2\la ca\ra^2\la da\ra^2 [\bar x x]}
\int{ d^d \ell \over \ell^2 A^2 B^2}
{[c|A|a\ra [d|B|a\ra [e|B|a\ra[e|A|a\ra[c|A|\bar x\ra [d|B|x\ra  [e|A| \bar x \ra    
\over(A+\bar x)^2}, 
\end{equation}

\begin{equation}\hskip-30pt
I^{[0]}_{x,\bar x:B_1}=
{[ex]\over [ab]^2\la ca\ra^2\la da\ra^2[\bar x x]}
\int{ d^d \ell \over \ell^2 A^2 B^2}
{[c|A|a\ra [d|B|a\ra [e|B|a\ra[e|A|a\ra[c|A|\bar x\ra [d|B|x\ra  [e|B| x\ra
\over(B-x)^2}, 
\end{equation}

\begin{equation}\hskip-30pt
I^{[0]}_{x,\bar x:A_2}=
{[e\bar x][xe]    [c\bar x]           \over [ab]^4\la ca\ra^2\la da\ra^2}
\int{ d^d \ell  \over \ell^2 A^2 B^2}
{[c|A|a\ra[d|B|a\ra [e|B|a\ra [e|A|a\ra \la \bar x|AB|x\ra  [d|A| \bar x\ra 
\over(A+\bar x)^2}, 
\end{equation} 
\begin{equation}\hskip-30pt
I^{[0]}_{x,\bar x:B_2}=
{[e\bar x][xe] [cx]    \over [ab]^4\la ca\ra^2\la da\ra^2}
\int{ d^d \ell  \over \ell^2 A^2 B^2}
{[c|A|a\ra [d|B|a\ra [e|B|a\ra [e|A|a\ra \la \bar x|AB|x\ra [d|B|x\ra
\over(B-x)^2} ,
\end{equation} 

\begin{equation}\hskip-30pt
I^{[0]}_{x,\bar x:A_3}=
{[e\bar x]^2[xe] [c\bar x][xd]                      \over [ab]^4\la ca\ra^2\la da\ra^2[\bar x x]}
\int{ d^d \ell  \over \ell^2 A^2 B^2}
{[c|A|\bar x\ra [d|B|x\ra [e|A|\bar x \ra  \la a|AB|a\ra^2 
\over(A+\bar x)^2}, 
\end{equation} 
\begin{equation}\hskip-30pt
I^{[0]}_{x,\bar x:B_3}=
{[ex]^2[\bar x e] [c\bar x][xd]                   \over [ab]^4\la ca\ra^2\la da\ra^2[\bar x x]}
\int{ d^d \ell  \over \ell^2 A^2 B^2}
{[c|A|\bar x\ra [d|B|x\ra  [e|B|x \ra \la a|AB|a\ra^2 
\over(B-x)^2}, 
\end{equation} 

\begin{equation}\hskip-30pt
I^{[0]}_{x,\bar x:A_4}=
{[ex]^2[e\bar x]^2 [c\bar x][xd] [\bar x|P|x\ra     \over [ab]^6\la ca\ra^2\la da\ra^2}
\int{ d^d \ell  \over \ell^2 A^2 B^2}
{[c|A|a\ra [d|B|a\ra  \la \bar x| AP|\bar x\ra \la a|AB|a\ra 
\over(A+\bar x)^2}, 
\end{equation} 
\begin{equation}\hskip-30pt
I^{[0]}_{x,\bar x:B_4}=
{[ex]^2 [e\bar x]^2 [c\bar x][xd] [x|P| x \ra                \over [ab]^6\la ca\ra^2\la da\ra^2}
\int{ d^d \ell  \over \ell^2 A^2 B^2}
{[c|A|a\ra [d|B|a\ra  \la x|BP|\bar x\ra \la a|AB|a\ra 
\over(B-x)^2}, 
\end{equation} 

\begin{equation}\hskip-30pt
I^{[0]}_{x,\bar x:A_5}=
{[ex]^2[e\bar x]^2 [c\bar x][xd] [\bar x|P|x\ra[x|P|\bar x\ra  [\bar x|P| a \ra \over [ab]^6\la ca\ra^2\la da\ra^2[\bar x x]}
\int{ d^d \ell  \over \ell^2 A^2 B^2}
{[c|A|a\ra[d|B|a\ra \la \bar x |AP|a\ra
\over(A+\bar x)^2},
\end{equation} 
\begin{equation}
\hskip-30pt
I^{[0]}_{x,\bar x:B_5}=
{[ex]^2[e\bar x]^2[c\bar x][xd] [\bar x|P|x\ra [x|P|\bar x\ra [x|P|a\ra  \over [ab]^6\la ca\ra^2\la da\ra^2[\bar x x]}
\int{ d^d \ell  \over \ell^2 A^2 B^2}
{[c|A|a\ra [d|B|a\ra \la x|BP|a\ra
\over(B-x)^2}   \; . 
\end{equation}

\section{Bubble Coefficients for $n$-point MHV amplitudes} 
\label{bubcoeapp}
In this appendix we present a  general formula for the bubble
coefficients of the $n$-graviton MHV scattering amplitude.   The bubble
coefficients vanish for $\NeqEight,6$ contributions.  The $\NeqFour$
contributions were given in ref.~\cite{Dunbar:2010fy} and the $\NeqOne$
coefficients in ref.~\cite{Dunbar:2011xw}.  We present the
remaining scalar contribution together with these in a
unified way using canonical forms~\cite{Dunbar:2009ax}. In the method of canonical
forms the bubble coefficients are obtained from unitarity by
decomposing   the product of tree amplitudes
appearing in a two-particle cut into canonical forms 
${\cal  F}_i$,
\begin{equation}
\sum M^{\tree}(-\ell_1, \cdots, \ell_2) \times
M^{\tree}(-\ell_2,\cdots,
 \ell_1) = \sum_i c_i {\cal F}_i ({\ell_j}),
\end{equation}
where the $c_i$ are coefficients independent of $\ell_j$.  The momentum
across the cut is $P=\ell_1-\ell_2$. We then use
substitution rules to replace the ${\cal F}_i ({\ell_j})$ by the
canonical forms $F_i(P)$ to obtain the coefficient of the bubble
integral $I_2(P^2)$ as
\begin{equation}
\sum_i c_i F_i(P)
\end{equation}
The general types of canonical forms we need for MHV amplitudes are
\begin{align}
{\cal H}_0^n &\equiv  \prod_{i=1}^n  [D_i|\ell_1|B_i\ra 
\\
{\cal H}_1^n &\equiv \prod_{i=1}^n  [D_i|\ell_1|B_i\ra  
{  \spa{\ell_1}.{B_{n+1}} \over \spa{\ell_1}.A } 
\\
{\cal H}_{1,1}^n &\equiv 
\prod_{i=1}^n  [D_i|\ell_1|B_i\ra   { \spa{\ell_1}.{B_{n+1}} \over
  \spa{\ell_1}.{A_1} } 
{ \spa{\ell_2}.{C_1} \over \spa{\ell_2}.{A_2} } 
\\
{\cal H}_{2x}^n &\equiv 
\prod_{i=1}^n  [D_i|\ell_1|B_i\ra   { \spa{\ell_1}.{B_{n+1}} \over
  \spa{\ell_1}.A } 
{ \spa{\ell_2}.{C_1} \over \spa{\ell_2}.A } 
\end{align}

Supersymmetric amplitudes require canonical forms
with lower values of $n$. In Yang-Mills theory we need
values $n\leq 2$ in general but only $n=0$ for supersymmetric contributions.
For gravity, scalar contributions require canonical forms up to $n= 4$ with only $n\leq 2$ for $\NeqOne$ and $n= 0$ for $\NeqFour$.
The expressions for the canonical forms with $n > 2$ were not given
previously. 

We first note some general results (not given previously)
\begin{align}
H^n_0[ \{B_i \}; \{D_i\}]  &
= {1\over (n+1)! } \sum_{P(B_i)}\prod_{i=1}^{n} [D_i|P|B_i\ra 
\notag  \\
H^n_1[A;  \{B_i\}; \{D_i\}]
&=  {1\over ( (n+1)!) ^2 }\sum_{P(B_i),P(D_i) } \sum_{r=0}^n { (P^2)^{n-r} \over  [A|P|A\ra^{n-r+1} }
\prod_{i=1}^{r} [D_i|P|B_i\ra \prod_{j=1}^{n-r} [D_i|A|B_j\ra [A|P|B_{n+1} \ra
\end{align}
$H^1_{2x}$ and $H_{2x}^2$ are given in ref~\cite{Dunbar:2009ax}.
The expressions for  $H^3_{2x}$ and $H_{2x}^4$ are
\begin{align}
& H_{2x}^3[ A; \{  B_1,B_2,B_3,B_4 \}; C_1 ; \{ D_1,D_2,D_3 \} ; P] 
= {1\over 576} \sum_{P(\{B_i\}),P(\{D_i\})} \Biggl(
\notag\\
&  { (P^2)^3\over  [A|P|A\ra^5 } \Biggl(
4 [D_1|A|B_1\ra  [D_2|A|B_2\ra [D_3|A|B_3\ra [A |P |B_4\ra
[A|P|C_1\ra
\notag\\
& \hskip 3.5 truecm - 4 [D_1|A|B_1\ra  [D_2|A|B_2\ra [D_3|A|C_1\ra [A |P |B_3\ra
[A|P|B_4\ra
\Biggr)
\notag\\
&  + { (P^2)^2\over  [A|P|A\ra^4 } \Biggl(
3 [D_1|A|B_1\ra [D_2|A|B_2\ra  [D_3|P|C_1\ra [A |P |B_3\ra [A|P|B_4\ra
\Biggr)
\notag\\
&  + { (P^2)\over  [A|P|A\ra^3 } \Biggl(
2 [D_1|A|B_1\ra  [D_2|P|B_2\ra [D_3|P|C_1\ra [A |P |B_3\ra [A|P|B_4\ra
\notag\\
& \hskip 3.5 truecm  
-4 [D_1|A|C_1\ra  [D_2|P|B_1\ra [D_3|P|B_2\ra [A |P |B_3\ra
[A|P|B_4\ra
\Biggr)
\notag\\
&  + { 1 \over  [A|P|A\ra^2 } \Biggl(
[D_1|P|B_1\ra [D_2|P|B_2\ra  [D_3|P|C_1\ra  [A |P |B_3\ra
[A|P|B_4\ra
\Biggr) \; \Biggr)
\end{align}
and
\begin{align}
 & H_{2x}^4[ A; \{ B_1,B_2,B_3,B_4,B_5 \} ; C_1 ; \{ D_1,D_2,D_3,D_4
 \}; P]  = {1\over 5!6!} \sum_{P(\{B_i\}),P(\{D_i\})} \Biggl(
\notag\\
&   { (P^2)^4\over [A|P|A\ra^6 } \Biggl(
12 [D_1|A|B_1\ra  [D_2|A|B_2\ra [D_3|A|B_3\ra [D_4 |A |B_4\ra [A |P |B_5\ra  [A|P|C_1\ra
\notag\\
& \hskip 3.5 truecm  
-2 [D_1|A|B_1\ra  [D_2|A|B_2\ra [D_3|A|B_3\ra [D_4 
|A |C_1\ra [A |P |B_4\ra  [A|P|B_5\ra
\Biggr)
\notag\\
&  + { (P^2)^3\over   [A|P|A\ra^5 } \Biggl(
6 [D_1|A|B_1\ra  [D_2|A|B_2\ra [D_3|A|B_3\ra [D_4 |P |C_1\ra [A |P |B_4\ra  [A|P|B_5\ra
\notag\\
& \hskip 3.5 truecm  
 -22[D_1|A|B_1\ra  [D_2|A|B_2\ra [D_3|A|C_1\ra [D_4 |P |B_3\ra [A |P  |B_4\ra  [A|P|B_5\ra
\Biggr)
\notag\\
&  + { (P^2)^2\over   [A|P|A\ra^4 } \Biggl(
 33[D_1|A|B_1\ra  [D_2|A|B_2\ra [D_3|P|B_3\ra [D_4 |P |B_4\ra [A |P |B_5\ra  [A|P|C_1\ra
\Biggr)
\notag\\
&  + { (P^2)\over  [A|P|A\ra^3 } \Biggl(
27 [D_1|A|B_1\ra  [D_2|P|B_2\ra [D_3|P|B_3\ra [D_4 |P |C_1\ra [A |P |B_4\ra  [A|P|B_5\ra
\notag\\
& \hskip 3.5 truecm  
-45 [D_1|A|C_1\ra  [D_2|P|B_1\ra [D_3|P|B_2\ra [D_4 |P |B_3\ra [A |P  |B_4\ra  [A|P|B_5\ra
\Biggr)
\notag\\
&  + { 1\over   [A|P|A\ra^2 } \Biggl(
-15  [D_1|P|B_1\ra  [D_2|P|B_2\ra [D_3|P|B_3\ra [D_4 |P |B_4\ra [A |P |B_5\ra  [A|P|C_1\ra
\notag\\
& \hskip 3.5 truecm  +
 21 [D_1|P|B_1\ra  [D_2|P|B_2\ra [D_3|P|B_3\ra [D_4 |P |C_1\ra [A |P |B_4\ra  [A|P|B_5\ra
\Biggr)
\;\Biggr)
\end{align}
We do not need to present $H_{1,1}^n$ separately since
\begin{align}
H_{1,1}^n[ A_1;A_2 ; B_1,\cdots & B_{n+1}; C_1;  D_{1},\cdots D_n ; P]
=H_2^n[ A_1,A_2 ; B_1,\cdots B_{n+1}, C_1; D_{1},\cdots D_n ; P]
\notag\\
+&{ \spa{ C_1}.{A_2} \spa{A_2}.{B_1} \over \spa{A_2}.{A_1} }
H_{2x}^{n-1} [ A_2 ; B_2,\cdots B_{n} ,B_{n+1};   P|D_1] ; D_{2},\cdots D_n ; P]
\notag \\
-&{ \spa{ C_1}.{A_2} \spa{B_1}.{A_1} \over \spa{A_2}.{A_1} }
H_{1,1}^{n-1}[ A_1;A_2 ; B_2,\cdots B_{n} , B_{n+1};  P|D_1] ; D_{2},\cdots D_n ; P]
\end{align}

We can now use these
canonical forms to provide expressions for 
the coefficients
of the bubble integrals $I_2(P^2)$. 
The bubble integral functions $I_2(P^2)$ will have vanishing
coefficients for the MHV amplitude unless the momentum $P$ (and hence
$-P$) contains exactly one negative helicity leg and at least one
positive helicity leg.  We can thus take $P$ to be of the form
$\{m_1^-,a_1^+,a_2^+,\cdots, a_{n_L}^+\}$ and the legs on the other
side of the cut to be $\{m_2^-,b_1^+,b_2^+,\cdots, b_{n_R}^+\}$.  Using the
canonical forms we find  the bubble coefficient
\begin{align}
& c(m_1,    \{a_i\}  ;m_2,\{b_j\} )
= { \spa{m_1}.{m_2}^M \over  (P^2)^{4-M} }
\sum_{P_L,P_R } 
C_{P_L} C_{P_R} \biggl( \notag \\
& 
\sum_{ x\in \{a_i\}\cup\{b_j\}-a_1} D_x { \spa{m_2}.{a_1}     \over \spa{b_1}.{a_1} }
H_{1,1}^{4-M}[ a_1;x; \{B_i\} ; m_1 ; \{D_i\} ; P] 
\notag \\
+&
\sum_{x\in \{a_i\}\cup\{b_j\}- b_1} D_x { \spa{m_2}.{b_1}     \over \spa{a_1}.{b_1} }
H_{1,1}^{4-M}[ b_1;x;\{B_i\} ; m_1 ;  \{D_i\} ; P] 
\notag \\
+ & 
D_{a_1}  { \spa{m_2}.{a_1}     \over \spa{b_1}.{a_1} }  H^{4-M}_{2x}[ a_1  ;\{B_i\} ; m_1 ;\{D_i\} ; P] 
+ 
D_{b_1} { \spa{m_2}.{b_1}     \over \spa{a_1}.{b_1} } H^{4-M}_{2x}[ b_1 ;\{B_i\} ; m_1  ;\{D_i\} ; P] 
\biggr) 
\label{allAform}
\end{align}
where $P_L$ and $P_R$ are permutations of the positive helicity legs $\{a_i\}$ and $\{b_i\}$ respectively.
The different cases are specified by  $M=2$ for the $\NeqFour$ matter multiplet, $M=1$ for the
$\NeqOne$ matter multiplet and $M=0$ for the scalar contribution. 
The  arguments are
\begin{align}
\{B_i\}^{\NeqFour}  =\{m_1\},
\{B_i\}^{\NeqOne}  =\{m_1,m_2,m_1\},
\{B_i\}^{\NeqZero}  =\{m_1,m_2,m_1,m_2,m_1\}
\notag \\
\{D_i\}^{\NeqFour}  =\emptyset,
\{D_i\}^{\NeqOne}  =\{ P|m_1\ra, P|m_2\ra \},
\{D_i\}^{\NeqZero}  =\{ P|m_1\ra, P|m_2\ra, P|m_1\ra, P|m_2\ra\}
\end{align}
and 
\begin{equation}
C_{P_L} ={ 1 \over \spa{n_L}.{m_1}
  \prod_{i=1}^{n_L-1} \spa{a_i}.{a_{i+1}} }   \; ,\;\;\; 
C_{P_R} ={ 1 \over \spa{n_R}.{m_2}
  \prod_{j=1}^{n_R-1} \spa{b_j}.{b_{j+1}} }   \; , 
\end{equation}

\begin{equation}
D_{x} = { \spa{m_2}.{x} \prod_{l=1}^{n_L-1} [ a_l|\tilde K_{l+1}|x\ra
  \prod_{k=1}^{n_R-1} [ b_k|\tilde K_{k+1}'|x\ra \over \prod_{y \neq
    x} \spa{x}.{y} } = { \prod_{l=1}^{n_L-1} [ a_l|\tilde K_{l+1}|x\ra
  \prod_{k=1}^{n_R} [ b_k|\tilde K_{k+1}' |x\ra \over
  \spb{b_{n_R}}.{m_2} \prod_{y \neq x} \spa{x}.{y} } 
\end{equation}
where
$\tilde K_p= k_{a_p}+\ldots k_{a_{n_L}} +k_{m_1}$ and $\tilde K_p'= k_{b_p}+\ldots k_{b_{n_R}} +k_{m_2}$.
The bubble coefficients satisfy, nontrivially, the IR relation $\sum_i c_i=0$~\cite{Dunbar:1995ed}.

\end{document}